\documentclass[
  journal=largetwo,
  manuscript=article-type,
  year=2022,
  volume=37,
]{cup-journal}

\usepackage{amsmath}
\usepackage{booktabs}
\usepackage{stfloats}
\usepackage{graphicx}
\usepackage{subfigure}
\usepackage{multirow}
\usepackage[colorlinks,bookmarksopen,bookmarksnumbered,citecolor=Blue, linkcolor=Red,urlcolor=MidnightBlue]{hyperref}
\DefineBibliographyStrings{english}{%
  bibliography = {abb},
}

\newcommand\apjs{Astrophysical Journal, Supplement}

\newcommand\apjl{Astrophysical Journal, Letters}

\newcommand\aap{Astronomy and Astrophysics}

\title{Estimating Stellar Parameters and Identifying Very Metal-poor Stars Using Convolutional Neural Networks for Low-resolution Spectra ($R\sim 200$)}

\author{Tianmin Wu}
\affiliation{School of Mathematics and Statistics, Shandong University, Weihai, 264209, Shandong, China}

\author{Yude Bu}
\email[Yude Bu]{buyude@sdu.edu.cn}
\affiliation{School of Mathematics and Statistics, Shandong University, Weihai, 264209, Shandong, China}

\author{Jianhang Xie}
\affiliation{School of Mathematics and Statistics, Shandong University, Weihai, 264209, Shandong, China}

\author{Junchao Liang}
\affiliation{School of Mathematics and Statistics, Shandong University, Weihai, 264209, Shandong, China}

\author{Wei Liu}
\affiliation{School of Mathematics and Statistics, Shandong University, Weihai, 264209, Shandong, China}

\author{Zhenping Yi}
\affiliation{School of Mechanical, Electrical \& Information Engineering, Shandong University, Weihai, 264209, Shandong, China}

\author{Xiaoming Kong}
\affiliation{School of Mechanical, Electrical \& Information Engineering, Shandong University, Weihai, 264209, Shandong, China}

\author{Meng Liu}
\affiliation{School of Mechanical, Electrical \& Information Engineering, Shandong University, Weihai, 264209, Shandong, China}


\begin{document}

\begin{abstract}
Very metal-poor (VMP, [Fe/H]<-2.0) stars offer a wealth of information on the nature and evolution of elemental production in the early galaxy and universe.
The upcoming China Space Station Telescope (CSST) will provide us with a large amount of spectroscopic data that may contain plenty of VMP stars, and thus it is crucial to determine the stellar atmospheric parameters ($T_{\rm{eff}}$, $\log$ g, and [Fe/H]) for low-resolution spectra similar to the CSST spectra ($R\sim 200$).
In this paper, a two-dimensional Convolutional Neural Network (CNN) model with three convolutional layers and two fully connected layers is constructed.
The principal aim of this work is to measure the ability of this model to estimate stellar parameters on low-resolution ($R\sim 200$) spectra and to identify VMP stars so that we can better search for VMP stars in the spectra observed by CSST. 
We mainly use 10,008 observed spectra of VMP stars from LAMOST DR3, and 16,638 spectra of common stars ([Fe/H]>-2.0) from LAMOST DR8 for the experiment and make comparisons. All spectra are reduced to $R\sim200$ to match the resolution of the CSST and are preprocessed and collapsed into two-dimensional spectra for input to the CNN model.
The results show that the MAE values are 99.40 K for $T_{\rm{eff}}$, 0.22 dex for $\log$ g, 0.14 dex for [Fe/H], and 0.26 dex for [C/Fe], respectively.
Besides, the CNN model efficiently identifies VMP stars with a precision of 94.77\%. The validation and practicality of this model are also tested on the MARCS synthetic spectra.
This paper powerfully demonstrates the effectiveness of the proposed CNN model in estimating stellar parameters for low-resolution spectra ($R\sim200$) and recognizing VMP stars that are of interest for stellar population and galactic evolution work.
\end{abstract}

\section{Introduction} 
\label{section:intro}
Very metal-poor (VMP, [Fe/H]<-2.0) stars are important relics of the Milky Way’s formation history, as the abundance of Li in these stars can provide an estimate of the baryon-to-photon ratio, and simulations predict that they are some of the oldest stars in the Galaxy \citep{2005ARA&A..43..531B}. By studying these stars, researchers can explore significant information about the Big Bang and the chemical and physical conditions of the first-generation stars in the universe \citep{2008ARA&A..46..241S, 2015ARA&A..53..631F, 2018ARNPS..68..237F}.

Owing to massive spectroscopic surveys, plenty of VMP stars have been detected to date.
The most noteworthy research should be the HK survey \citep{1985AJ.....90.2089B, 2001AAS...199.9108R,2003AAS...20311213R} and the Hamburg/ESO Survey  \citep{2000A&A...358...77W,2002AJ....124..470C,2006ApJ...652.1585F,2008A&A...484..721C}, leading to the discovery of approximately 2,000 VMP stars. 
In the follow-up SDSS/SEGUE program, over 16,000 VMP stars were observed from medium-resolution spectra, extremely expanding the VMP stars database \citep{e851d5c21b744415902a89fd7af69ec9}. 
Additionally, among the VMP candidates with photometric metallicity abundances greater than or equal to -2.0 obtained from SkyMapper DR1.1, researchers found nearly 2,500 stars with metallicity less than -2.0 with the help of the follow-up low-resolution ($R\sim3000$) spectroscopic research \citep{2019MNRAS.489.5900D}. The dedication of other large sky surveys such as APOGEE \citep{2013ApJ...767L...9G} and RAVE \citep{matijevivc2017very, 2022ApJ...927...13Z} also enhanced our study of VMP stars.
In 2009, the innovative Large Sky Area Multi-Object Fiber Spectroscopic Telescope (LAMOST) was completed in China. Due to LAMOST's deeper observing depth and its ability to acquire large multi-fiber samples, the observation of medium or low-resolution spectra ($R\sim1000$ or 2000) from LAMOST has made significant contributions to the search for VMP stars \citep{2010New}.
Via the LAMOST DR1 dataset, \citet{2015ApJ...798..110L} reported the early results about nearly 100 very metal-poor star candidates. High-resolution spectroscopic identification of eight of these candidates shows that all of them are VMP stars. In the subsequent DR3 dataset released by LAMOST, \citet{2018ApJS..238...16L} made use of synthetic spectra with line indices and made comparisons to the observed spectra to find the best-fitting stellar parameters, successfully confirmed 10,008 VMP stars in the dataset. \citet{2022ApJS..259...51W} crossed LAMOST DR8 low-resolution spectroscopic data with PASTEL labels and APOGEE DR16 data, applying a neural network approach, to greatly improve the estimation of [Fe/H] and obtain a catalog of 26,868 VMP stellar candidates.

Metal-poor stars tend to contain higher than average levels of carbon. If the carbon abundance ([C/Fe]) of the metal-poor  ([Fe/H]<-1.0) stars is larger than +1.0, it is called the Carbon Enhanced Metal-poor (CEMP) stars \citep{2005ARA&A..43..531B}. This threshold for classifying CEMP stars has been updated to [C/Fe]>+0.7 \citep{2007ApJ...655..492A}. Measuring the carbon enhancement of metal-poor stars discovered from large-scale surveys is conducive to deriving CEMP stars, which are of vital importance for understanding the relationship between astrophysical s-process and carbon enhancement \citep{2005NuPhA.758..312M} and the nature of first-generation stellar evolution, especially for VMP stars \citep{2011hst..prop12554B}. \citet{2006ApJ...652L..37L} analyzed 94 VMP stars obtained by the Hamburg/ESO R-process Enhanced Star (HERES) survey \citep{barklem2005hamburg} and found $21\pm2\%$ of VMP stars with [C/Fe] abundances above +1.0, which can be classified as CEMP stars. Although the frequency of CEMP stars in VMP giants ($9\pm2\%$) derived by \citet{2006ApJ...652.1585F} is not good enough, they discovered clear evidence about the increase of the proportion of carbon enrichment in metal-poor stars with decreasing metallicity, which is valuable for studying CEMP stars. A more productive method based on matching spectra with a custom grid of synthetic spectra was presented by \citet{2013AJ....146..132L}, which could be used to obtain the fractions of CEMP stars in metal-poor stars from a large sample of SDSS/SEGUE low-resolution ($R\sim2000$) spectra with precision over 0.35 dex. Utilizing the most extensive high-resolution samples from a variety of literature at that time, \citet{2014ApJ...797...21P} improved the frequency of CEMP stars in metal-poor stars and derived that 20\% of VMP stars have [C/Fe] abundance greater than +0.7. \citet{2022MNRAS.515.4082A} collected the results of CEMP studies over 25 years and compared them, finding significant differences in the distribution of CEMP stars in the giant star samples and suggesting some constructive advice for the follow-up studies.

In order to identify VMP and CEMP stars, the metallicity and [C/Fe] abundance should first be determined. Many methods have been proposed to extract these stellar labels from the massive spectroscopic data obtained from large sky surveys. Depending on the medium-resolution spectra of SDSS-\textrm{i}and SDSS-\textrm{ii}/SEGUE, the SEGUE Stellar Parameter Pipeline (SSPP) was first raised by \citet{2008AJ....136.2022L}. They verified the accuracy and quantified error of this method and illustrated its effectiveness in measuring stellar parameters for large sky observations. Some automated approaches, such as ULySS \citep{2009A&A...501.1269K} and iSpec \citep{2014A&A...569A.111B}, are complete software packages for stellar spectral analysis and parameter estimation. The LAMOST stellar parameter pipeline (LASP) can also be used to derive fundamental stellar parameters automatically \citep{2014IAUS..306..340W}. In the cases of difficulties in modeling due to the acquisition of an excessive amount of spectral data, \citet{2015ApJ...808...16N} developed The Cannon, a data-driven method that offers researchers the opportunity to obtain stellar parameters and chemical abundances from lower signal-to-noise (S/N) spectra with essentially no degradation in their accuracy. \citet{2019MNRAS.483.3255L} created a Python package named astroNN, combining the artificial neural network (ANN) with Bayesian dropout variational inference, which can effectively analyze high-resolution spectra to calculate stellar parameters. The researchers propose a general method Payne for fitting stellar spectra and physical models and simultaneously predicting stellar labels \citep{2019ApJ...879...69T}. They tested the validation of the method on the APOGEE DR14 high-resolution spectroscopic dataset. Another method called Stellar Parameters and Chemical Abundances network (SPCANet) was provided by \citet{2020ApJ...891...23W}, which is based on Convolutional Neural Network (CNN) and applied explicitly to the LAMOST medium-resolution spectra dataset. The semi-parallel structure with two branches of convolutional layers is considered the most particular part of this network. \citet{2022AJ....163..153L} also considered a deep learning algorithm called Light gradient boosting machine (LightGBM) to measure stellar parameters after extracting stellar characteristics using principal component analysis (PCA). This method can handle the instability of the model and improve the estimation accuracy. Nevertheless, as mentioned above, most approaches for measuring stellar parameters are based on high-resolution spectra or spectra with $R\sim2000$ down to 1000 and there is a paucity of studies in spectra of $R\sim200$. Therefore, our study would be an important supplement to the above methods, providing additional algorithmic options.

The Chinese Space Station Telescope (CSST), also known as the Chinese Survey Space Telescope, is a large-scale optical telescope with a diameter of two meters \citep{zhan2021wide}. It integrates high-quality performance with a large field of view and high image quality, as well as the ability to maintain and upgrade on-orbit. It can cover an observation area of 17,500 $deg^2$ and a wide wavelength range (255-1,000 nm by three bands GU, GV, and GI) and is expected to launch into low-Earth orbit around 2024. The main goal of the CSST is to carry out seamless spectral surveys and multi-band imaging and to provide high-quality low-resolution ($R>200$) seamless spectra for hundreds of millions of celestial objects \citep{2021RAA....21...74Y, 2021RAA....21...92S}. The development of new methods for estimating stellar parameters of the spectra of $R\sim200$ will be of great help in studying the low-resolution spectra obtained by the CSST that may cover plenty of VMP stars. In this context, we design a two-dimensional CNN model that includes three convolutional layers and two fully connected layers. We use the spectral data obtained from LAMOST and reduce its resolution to $R\sim200$ to validate our model. MARCS synthetic spectra and other machine learning methods are also used to test whether our model has higher accuracy.

The paper consists of five parts. The data we use is illustrated in Section \ref{section:data}. Then we elaborate our CNN model in Section \ref{section:method}. Section \ref{section:result} presents the experiments and results. Section \ref{section:discussion} discusses the comparison between CNN models and other machine learning algorithms, and brief conclusions can be seen in Section \ref{section:conclusion}.

\section{Data} 
\label{section:data}
Two types of data are used in this paper, the LAMOST database and MARCS synthetic spectra.
\subsection{LAMOST} 
\label{subsection:lamost}
\subsubsection{Data introduction} 
\label{subsubsection:dataintro}
The Large Sky Area Multi-Objective Fiber Spectroscopic Astronomical Telescope (LAMOST), also known as Guo Shoujing Telescope, located at the Xinglong station of the National Astronomical Observatory, is a transverse north-south Reflecting Schmidt Telescope \citep{2012RAA....12.1197C}. As a representative among spectroscopic survey telescopes, LAMOST applies thin mirror active optics and spliced mirror active optics technology, which ingeniously realizes an optical telescope with both large fields of view and a large aperture. Due to its 4-meter diameter, faint celestial bodies of magnitude 20.5 can be observed, and 4,000 optical fibers can be placed on a 5-degree field of view. Another breakthrough technology of LAMOST is the parallel controllable fiber positioning technology, which can precisely locate 4,000 observation targets and obtain the spectra of 4,000 objects simultaneously. This is also an internationally leading technology that dramatically improves the efficiency of the census of celestial objects. LAMOST has enabled people to reach an unprecedented level of understanding of the Milky Way galaxy and has also promoted the leap-forward development of technology for building large-aperture astronomical telescopes in China \citep{2022MNRAS.514.4588L}. So far, LAMOST has released the eighth version of the data to the public, containing 10,633,515 low-resolution ($R\sim 1800$) spectra of 10,336,752 stars, 224,702 galaxies, and 72,061 quasars. The basic stellar atmospheric parameters of A, F, G, and K types of stars are automatically derived from the LAMOST stellar parameter pipeline (LASP) \citep{2014IAUS..306..340W}. The determination of M-type stars is developed by LASP-M \citep{2021RAA....21..202D}.

There are 26,646 stars in the dataset, including 10,008 VMP stars ([Fe/H]<-2.0) and 16,638 common stars ([Fe/H]>-2.0). All spectral data are from LAMOST low-resolution observation ($R\sim 1800$). The VMP stars catalog was derived by \citet{2018ApJS..238...16L} from LAMOST DR3. The reason why we chose to use this dataset is that it still provides researchers with the largest pool of bright and accurate VMP candidates to date. The stellar parameters of these VMP stars were determined by line indices and by comparison with a grid of synthetic spectra, with metallicity ranging from -4.5 dex to -2.0 dex. They also present that the CEMP stars can be detected by G1 and EGP \citep{2011AJ....142..188P} line indices sensitive to carbon enhancement. Based on the criterion of G1 > 4.0 $\mathring{A}$ and EGP > -0.7mag, 636 CEMP stars were distinguished and labeled from the VMP stars catalog. Since \citet{2018ApJS..238...16L} did not provide [C/Fe] values for these VMP stars, we crossed these 10,008 VMP stars with the stellar parameter catalog offered by \citet{2020ApJ...891...39Y} and found the [C/Fe] values for 8,117 of the VMP stars, which were obtained with high accuracy from the SEGUE Stellar Parameter Pipeline (SSPP). These 8,117 VMP stars were used in Section \ref{subsection:c} to verify the ability of the proposed CNN model to estimate [C/Fe] values.

\begin{figure}[htbp]
		\centering
			\begin{minipage}{\linewidth}
				\centering
				\includegraphics[width=4in, height=3in]{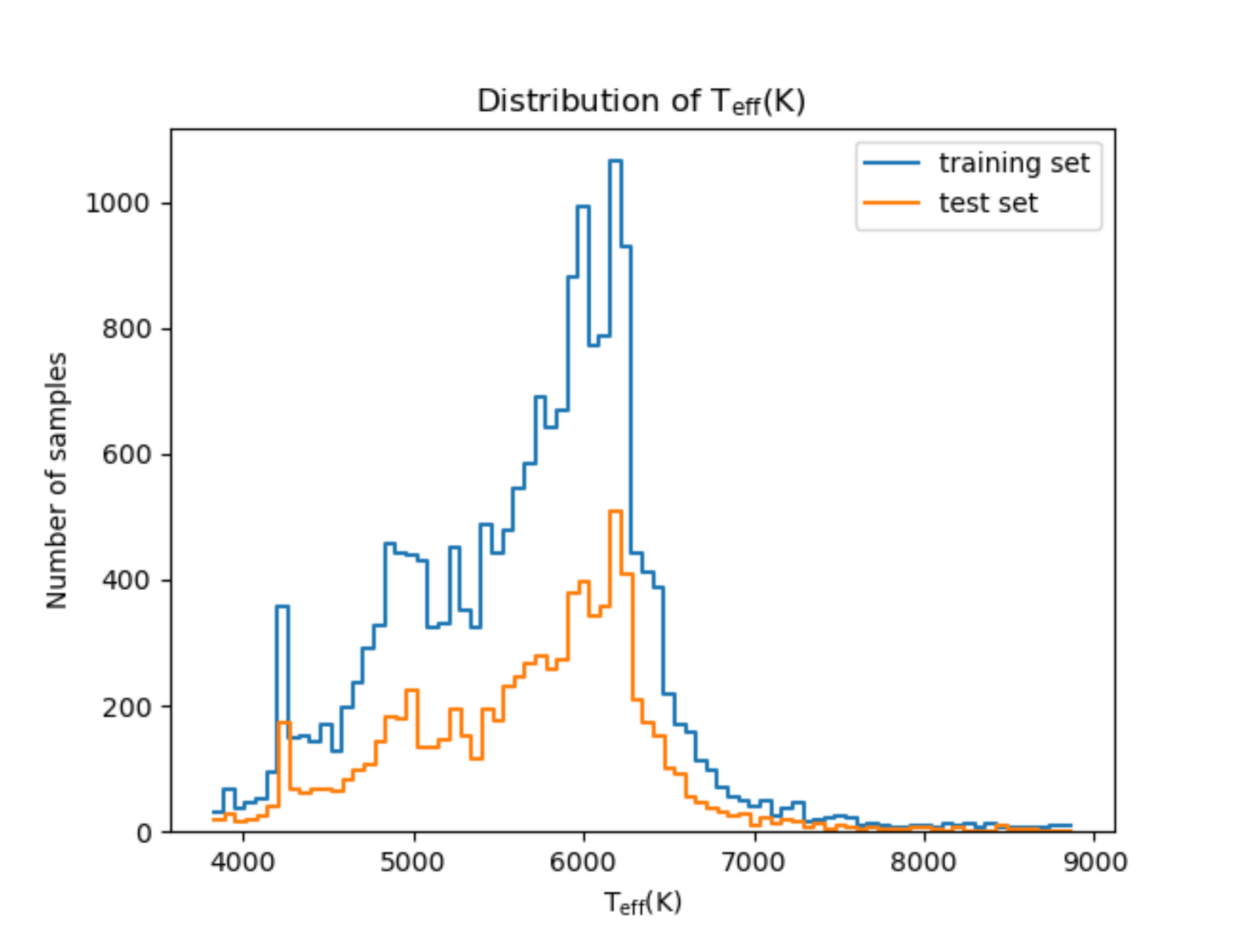}
			\end{minipage}%
		\qquad
			\begin{minipage}{\linewidth}
				\centering
				\includegraphics[width=4in, height=3in]{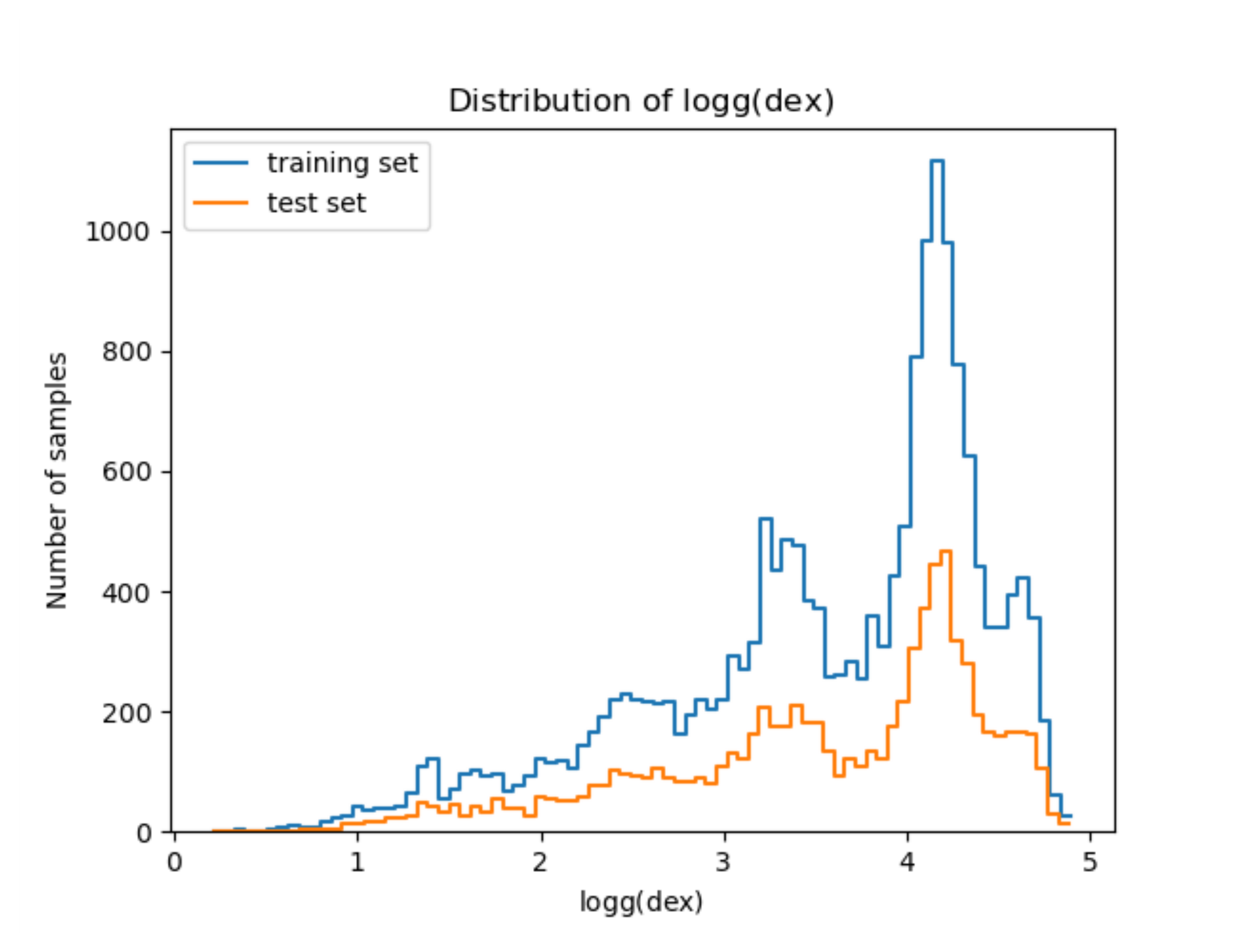}
			\end{minipage}%
		\qquad
			\begin{minipage}{\linewidth}
				\centering
				\includegraphics[width=4in, height=3in]{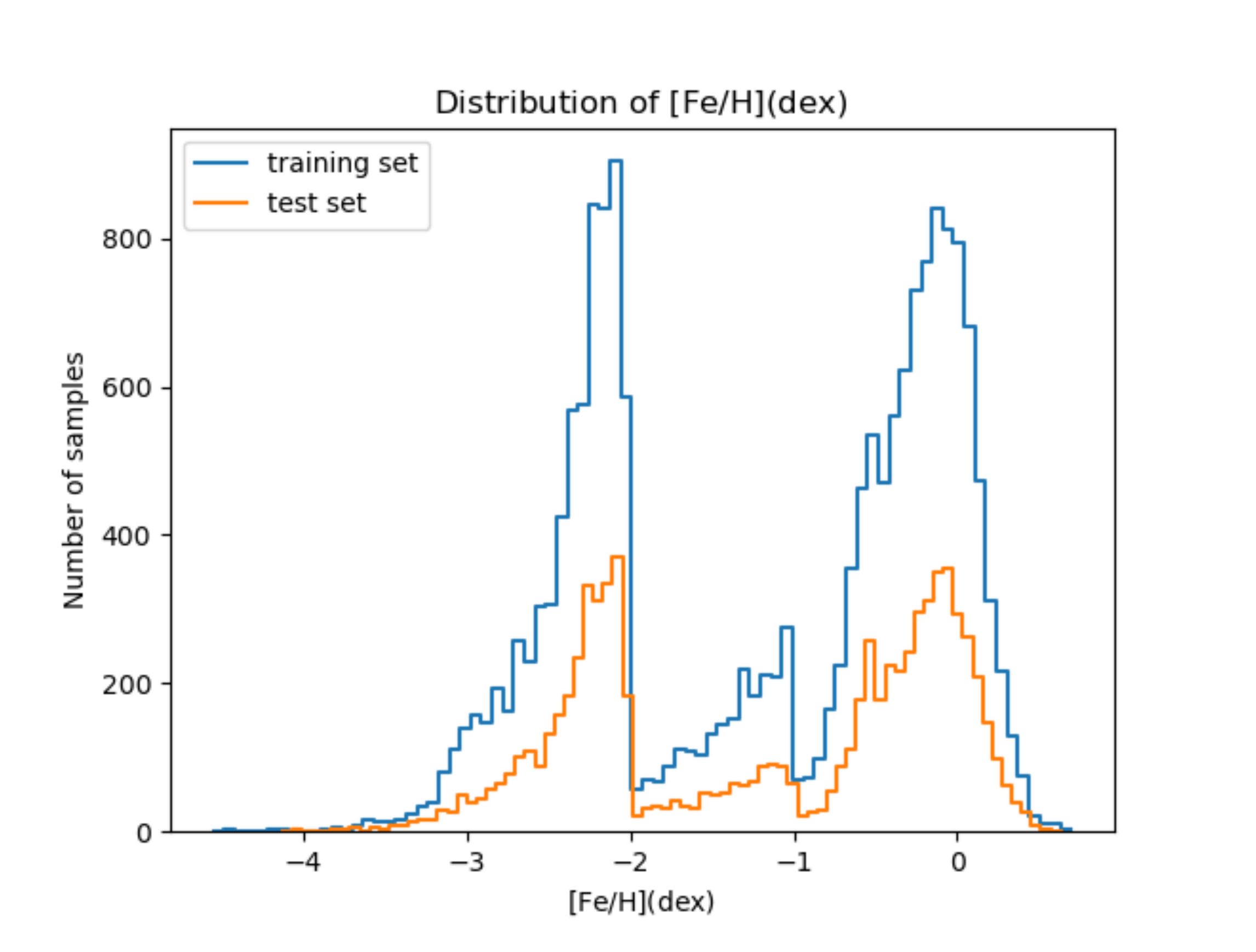}
			\end{minipage}%
		\caption{The distribution of the fundamental atmospheric stellar parameters on the training set and test set.}
		\label{fig:parameters}
\end{figure}

Since the approach that \citet{2018ApJS..238...16L} used is not very suitable for defining stellar parameters of common stars, other 16,638 common stars come from the recently published LAMOST DR8 dataset, where the stellar parameters were determined by LASP. We randomly selected data with [Fe/H]>-2.0 and minor uncertainties, which have a signal-to-noise ratio larger than ten at g-band, to ensure that the accuracy of the stellar parameters of this dataset is comparable to that of the VMP stars dataset. The parameters of these 26,646 stars range from 3824.88 K<$T_{\rm{eff}}$ <8866.15 K, 0.213 dex<$\log$ g<4.897 dex, and -4.55 dex<[Fe/H]<0.699 dex. The errors of the paramaters range from 0 K<$\sigma(T_{\rm{eff}})$ <399 K, 0 dex<$\sigma(\log g)$<0.94 dex, and 0 dex<$\sigma([Fe/H])$<0.4 dex. Figure \ref{fig:parameters} depicts more clearly the distribution of the parameters in the training and test sets.

\subsubsection{Data pre-processing} 
\begin{figure*}[htbp]
		\centering
		\includegraphics[width=7in, height=2.5in]{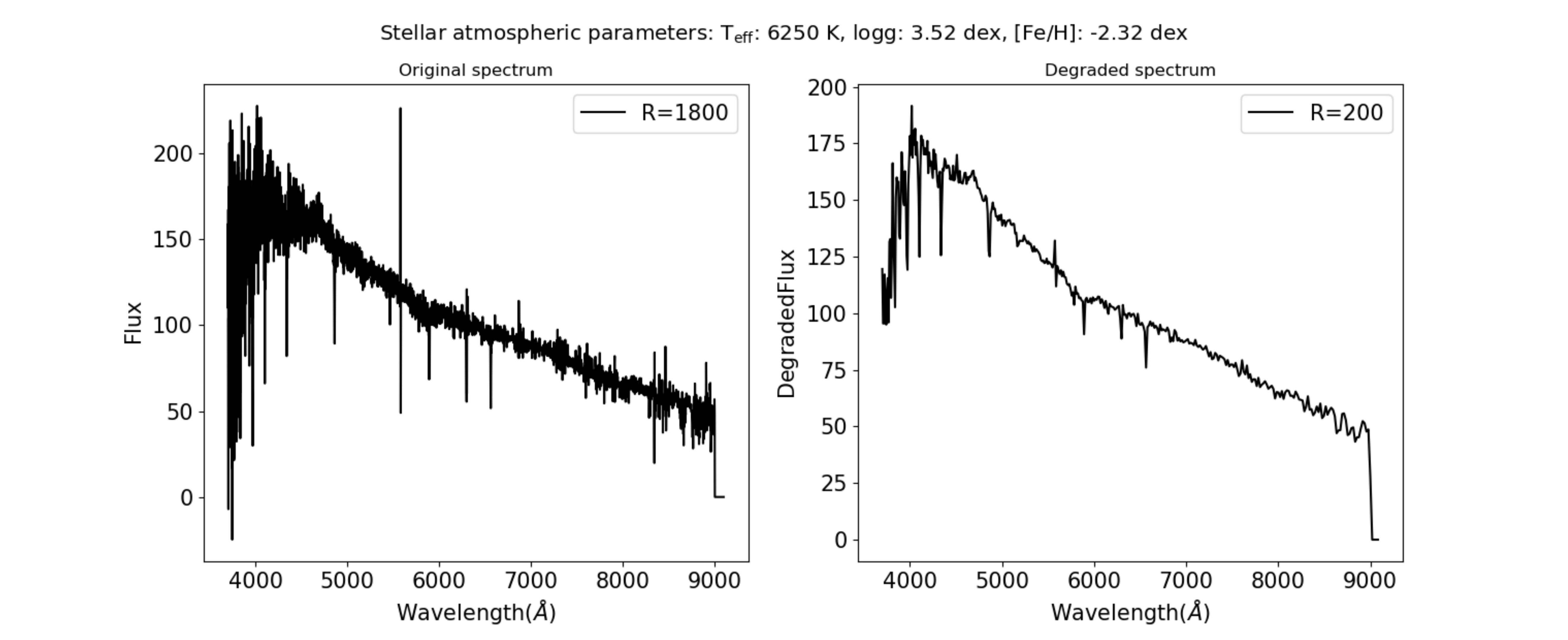}
		\caption{The demonstration plots of spectra with resolution $R\sim 1800$ and $R\sim 200$.}
       \label{fig:spectrum}
\end{figure*}

To begin with, we reduce the resolution of the spectra from $R\sim1800$ to $R\sim200$ with 391 feature points to simulate the low-resolution spectra acquired by CSST. The Coronagraph library provided in Python with the noise\_routines.construct\_lam() and downbin\_spec() functions can bring the data down to the resolution we need and output the degraded flux. Since the flux range is inconsistent for each spectrum and there is a lot of noise at both ends of the spectrum, we interpolate the flux data to 4000 to 8095$\mathring{A}$ to obtain spectra with 410 feature points. This reduces the effect of noise, and the more sample points obtained help the algorithm to be more effective.
On this basis, the flux values are then normalized by a linear function (Min-Max scaling), which can achieve equal scaling of the original data to convert the flux to the range of [0, 1], as follows. 
\begin{equation}
	Flux_{norm}=\frac{Flux-Flux_{min}}{Flux_{max}-Flux_{min}},
\end{equation}
As a result, the final spectral data can be obtained. The demonstration plots of  the spectra with resolutions of $R\sim1800$ and $R\sim200$ are shown in Figure \ref{fig:spectrum}.

\subsection{MARCS synthetic spectra} \label{marcs} 
Another dataset used in this paper comes from the MARCS synthetic spectra that \citet{2008A&A...486..951G} created, which is a grid of about $10^{4}$ model atmospheres with nearly 52,000 stellar spectra containing F, G, and K types of stars. This grid of one-dimensional LTE model atmospheres can be combined with atomic and molecular spectral line data and software to generate stellar spectra, which has been widely used in a variety of studies. \citet{2014AJ....147..136R} applied MARCS model atmospheres in order to process standard LTE abundance analysis, and statistical corrections were used to minimize systematic differences, which made contributions to measuring detailed abundances of 313 VMP stars. \citet{2018AcASn..59...35L} measured $[\alpha/Fe]$ using the Haar wavelet and LASSO methods and verified its validity in the MARCS stellar spectral library. Other studies \citep{2019A&A...627A.177R, 10.1093/mnras/stab2996, 2022A&A...662A.120S} also illustrate the practicality of MARCS synthetic spectra. Therefore, the wide applicability of our model can be further seen by using synthetic spectra to conduct experiments.

Here we select 9,644 of the MARCS synthetic models for the experiments to further verify the efficiency of the CNN model. The range of the stellar parameters is 2500 K<$T_{\rm{eff}}$ <8000 K, -0.5 dex<$\log$ g<5.5 dex, -5 dex<[Fe/H]<-1 dex and the step sizes of the parameter distributions are 2500 K for $T_{\rm{eff}}$, 0.5 dex for $\log$ g, and 0.25 dex for [Fe/H], severally. After the interpolation and normalization, the one-dimensional spectral data containing 746 features is folded to a $28\times28$ two-dimensional matrix and used as the input to the CNN model. Section \ref{subsection:exp2} shows the specific results of estimating stellar parameters of the MARCS synthetic spectra using the CNN model.

\section{Methodology} \label{section:method}
\subsection{Introduction to the convolutional neural network (CNN)}
Deep learning has been widely used in various fields in recent years. The concept originated from the study of artificial neural networks and was proposed by Hinton et al. in 2006. This paper utilizes a CNN model to test its performance in estimating the stellar atmospheric parameters of the low-resolution ($R\sim200$) spectra and better identify the VMP stars in the universe, laying a foundation for us to explore the deepest mysteries of the universe.

Convolutional neural networks, presented by \citet{726791}, are the first actual multi-layer structure learning algorithm that uses spatial relative relationships to reduce the number of parameters to improve training performance. It is a machine learning model under deep supervised learning. On the basis of the original multi-layer neural network, a feature part is added; that is, a convolutional layer and a pooling layer (dimension reduction layer) are added before the fully connected layer, and the network selects the features by itself. It is a deep feedforward neural network and is widely used in supervised learning problems of image processing and natural language processing, such as computer vision, semantic segmentation, object recognition, etc. The more classic CNNs are AlexNet, LeNet-5, VGG, etc. Generally speaking, the input data of CNN is a two-dimensional image in the form of RGB. A complete CNN model must include convolutional layers, non-linear activation functions, pooling layers, and fully connected layers. Moreover, sometimes, we need normalization layers, dropout layers, etc., to prevent gradient explosion, gradient disappearance, and overfitting problems.
\begin{enumerate}
    \item Convolutional layer: 

The convolutional layer is the core layer for establishing the CNN model, which can act as a filter and reduce the number of parameters.
Let $f(x)$ and $g(x)$ be the integrable functions, one-dimensional convolution is defined as
\begin{equation}
	(f\ast g)(t)=\int f(t)g(t-\tau)d\tau,
\end{equation}
	 the discrete form is
\begin{equation}
	(f\ast g)(t)=\sum_{\tau=-\infty}^{\infty} f(t)g(t-\tau).
\end{equation}
In convolutional neural networks, the discrete form of two-dimensional convolution is usually used.
Given a figure $\textbf{X}\in \textbf{R}^{M\times N}$  and a convolutional kernel 
$\textbf{W}\in\textbf{R}^{U\times V}$. 
In general, U<M, V<N. The convolution between them can be denoted as 
\begin{eqnarray}
		\mathbf{Y}=\mathbf{W}*\mathbf{X},\\
		y_{ij}=\sum_{u=1}^{U}\sum_{v=1}^{V}w_{uv}x_{i-u+1,j-v+1}.
\end{eqnarray}

The convolutional layer extracts features from local regions, and different convolutional kernels are equivalent to various feature extractors. Based on the standard definition of convolution, strides and zero padding of the convolutional kernel can also be introduced to increase the diversity of convolution. Strides refer to the number of steps each convolutional kernel moves when performing a convolutional operation. Set stride=$k$, which means convolving $k$ rows and $k$ columns from left to right and from top to bottom. Zero padding represents adding zeros to the outer side of the image. Setting padding=$d$, which means supplementing $d$ layers of zeros around the input vector. Zero padding allows us to obtain more detailed feature information and control the network structure. After the convolutional layer, an activation function is usually added as a non-linear factor, which will deal with problems that cannot be solved by linear models and enhance the ability of the network to interpret the model. The commonly used activation functions include ReLU, Leaky, sigmoid function, etc.

A simple example of the convolutional process is shown in Figure \ref{convolution}. A two-dimensional input array ($3\times3$) performs a mutual correlation operation with a two-dimensional convolutional kernel array ($2\times2$), resulting in a two-dimensional array ($2\times2$). The convolution kernel slides over the input array from left to right and top to bottom.

\begin{figure}[htbp]
		\centering
		\includegraphics[width=0.9\linewidth]{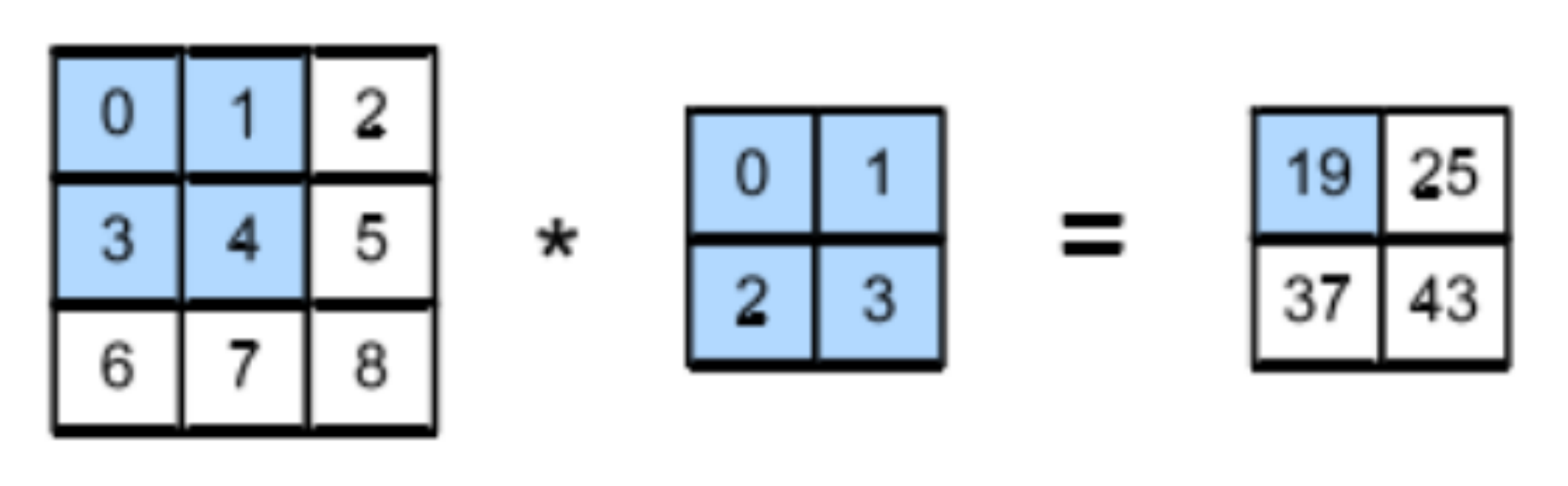}
		\caption{Convolutional Process}
       \label{convolution}
\end{figure}

    \item Pooling layer: 
The pooling layer, also named the subsampling layer, is designed to reduce the number of features in the network, thereby reducing the number of parameters and avoiding overfitting.
Max-pooling (giving the maximum value within the neighborhood) and mean-pooling (giving the average within the neighborhood) are two widely used pooling functions. Local translation invariance is an essential property of pooling layers, which indicates that pooling is approximately invariant in its representation of the input when a small number of translations are performed on the input.
    \item Fully-connected layer: 
The purpose of the fully-connected layer is to connect the result of the last pooling layer to the output node and map the feature representation learned by the network to the label space of the sample. It acts as a "classifier" on the network. It should be noted that when encoding the model, the last pooling layer needs to be flattened first to make it a one-dimensional vector before connecting to the fully connected layer.
\end{enumerate}
The feed-forward of the convolutional neural network is to extract high-level semantic information from the input layer step by step through a series of operations such as convolution, pooling, and mapping of nonlinear activation functions, and ultimately formalize the target task (regression or classification, etc.) into an objective function through a fully connected layer to output the predicted value. The task is to train the CNN model, that is, by calculating the error or loss between the predicted value and the true value, back-forward the error or loss layer by layer with the help of backpropagation, update the parameters and repeat this process until the model converges.

\subsection{The structure of the proposed CNN model}
\begin{figure}[htbp]
		\centering
		\includegraphics[width=0.9\linewidth]{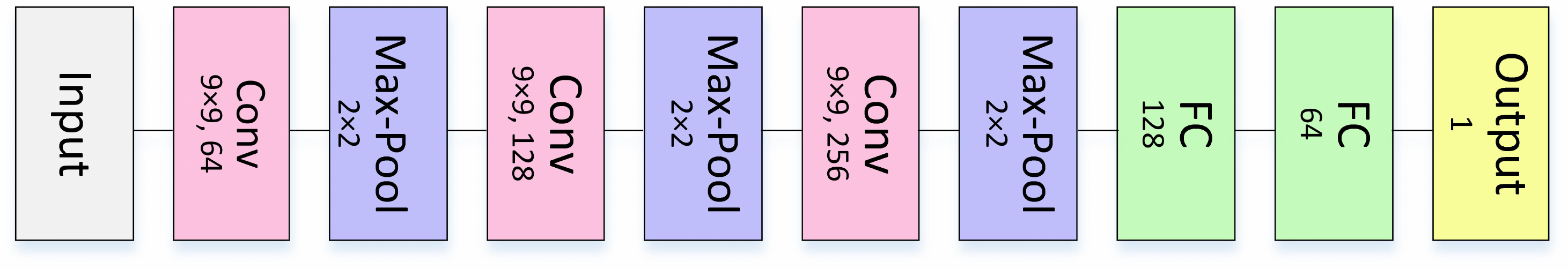}
		\caption{The structure of the CNN model used in the experiments.}
       \label{fig:cnn}
\end{figure}

Since Convolutional Neural Networks are very capable of extracting features from images, we collapse the one-dimensional spectral data containing 410 features into $21\times 21$(441) two-dimensional spectral data, which is fed into the convolutional neural network as an image, and the remaining 31(441-410) features can be filled with zeros.
After parameter tuning of the model, our final CNN model consists of six trainable layers. The convolutional layers filter the processed 2D spectra using a filter of size $9\times9$, and the input is filled with zero space on the boundary so that the size of the output layer of the convolution is equal to the size of the input layer. The $9\times9$ convolution kernel can acquire a larger field of perception and therefore can capture more characteristics. Then the convolutional layers are followed by a max-pooling layer of size $2\times2$ with a step size set to two. Max-pooling is better at preserving texture features than mean-pooling. Afterwards, two fully connected layers with 128 and 64 channels are added to combine the features previously extracted by the model. To prevent overfitting, a dropout layer can be set after each fully connected layer with a value of 0.2 to avoid the over-regularization of the model. Batch normalization (BN) and ReLU activation function layers are added between each layer to reduce overfitting and enhance the expressiveness of the model.
The final output layer is the predicted values of the stellar atmospheric parameters derived from the model. 
The detailed parameter settings of the model are given in Figure \ref{fig:cnn}. The kernels for each convolutional layer are 64, 128, and 256, severally.

\section{Results} \label{section:result}
In this section, we use spectral data described in Section \ref{section:data} to test the effectiveness of the proposed CNN model in estimating the stellar atmospheric parameters and classifying VMP stars. We divide the dataset into a training set and a test set in the ratio of 7:3 to train the model and measure its accuracy.
The training process is performed in batches, with the size of each batch set to 128. The Adam algorithm, with an initial learning rate set to 0.001, is chosen to be the optimizer for model training, which is an extension of the stochastic gradient descent method. \citet{kingma2014adam} suggested that the advantage of the Adam optimizer is that it can speed up convergence by adapting the learning rate, making it well-suited for deep learning problems.
A total of 1,000 training epochs are set, and the L1loss (MAE, Equation \ref{eq:mae}) is used as the loss function. 
An early stopping mechanism is set when the loss function no longer decreases beyond 250 epochs, which can effectively prevent the model from overfitting.

Three evaluation indicators for validating the effect of the model are shown below.
Suppose that $N$ is the number of samples contained in the test set, $y$ denotes the true values, and $\hat y$ denotes the predicted values derived by the proposed model. Let $e_{i}$ be $y_{i}-\hat y_{i}$, and  $\bar e_{i}$ be the average value of $e_{i}$
\begin{enumerate}
 \item Mean absolute error (MAE):
  MAE is a loss function used in regression models, which can express the fitting ability of the model more intuitively.
  \begin{equation}
  MAE(y,\hat y)=\frac{1}{N}\sum_{i=0}^{N-1}|y_{i}-\hat y_{i}|.
    \label{eq:mae}
  \end{equation}

 \item Standard deviation (STD):
Standard deviation reflects the degree of dispersion of a dataset.
  \begin{equation}
  STD(y,\hat y)=\sqrt {\frac{1}{N}\sum_{i=1}^{N}(e_{i}-\bar e_{i})^2}.
  \end{equation}
 \item Median value of error (M):
 The median value of the error can provide a better measure of the outliers in the dataset.
  \begin{equation}
  M(y,\hat y)=median(e_{i}).
  \end{equation}
\end{enumerate}

\subsection{Estimating stellar parameters and classifying VMP stars using the LAMOST dataset of 26,646 stars}\label{subsection:exp1}
We first conduct experiments using a total of 26,646 spectra, including VMP and common stars, with a resolution of 200 to test the ability of the proposed CNN model to estimate stellar parameters on low-resolution spectra. The dataset is divided with a ratio of 7:3 into a training set containing 18,652 stars and a test set containing 7,994 stars. The training process is performed on $T_{\rm{eff}}$, $\log$ g and [Fe/H] respectively. The prediction results obtained on the test set are displayed in Table \ref{tab:table1}. For $T_{\rm{eff}}$, MAE=99.40 K, STD=183.33 K, M=-0.49 K; for $\log$ g, MAE=0.22 dex, STD=0.35 dex, M=-0.02 dex; and for [Fe/H], MAE=0.14 dex, STD=0.26 dex, M=0.01 dex. 

\begin{table}[htbp]
  \centering
  \caption{The prediction results of the three fundamental atmospheric parameters  on test set including 7,994  stars using the proposed CNN model.}
  \label{tab:table1}
  \setlength{\tabcolsep}{5mm}
  \begin{tabular}{cccc}
\toprule
Parameter           & MAE    & STD    & M    \\ \midrule
$T_{\rm{eff}}$ (K)  & 99.40  & 183.33 & -0.49 \\
$\log$ g(dex)       & 0.22   & 0.35   & -0.02   \\
{[Fe/H]}(dex)       & 0.14   & 0.26   & 0.01  \\ 
\bottomrule
  \end{tabular}
\end{table}

\begin{figure*}[htbp]
		\centering
		\subfigure{
			\begin{minipage}[t]{0.49\linewidth}
				\centering
				\includegraphics[width=3.5in, height=2.6in]{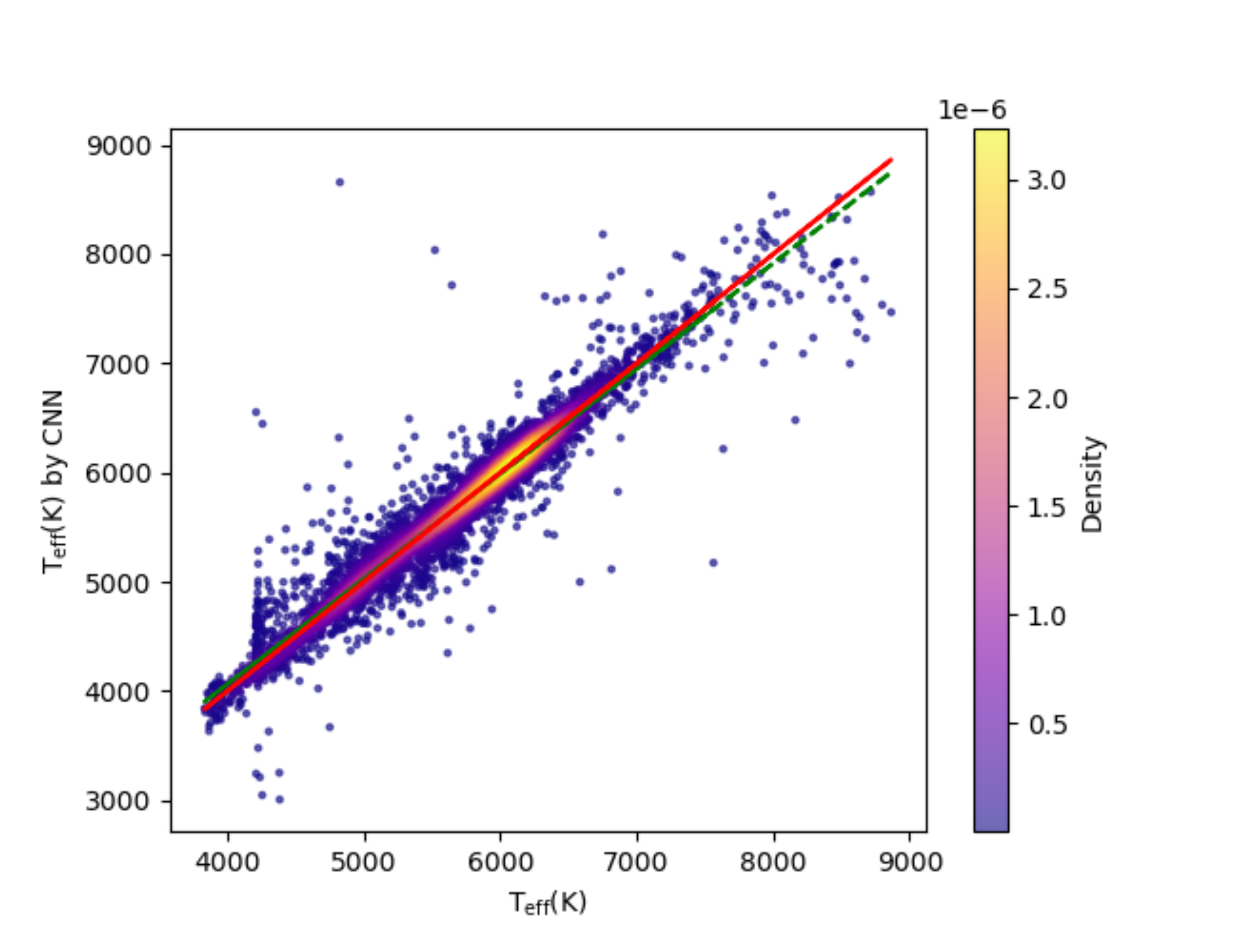}
				\centering
			\end{minipage}%
		}\hspace{-3mm}%
		\subfigure{
			\begin{minipage}[t]{0.49\linewidth}
				\centering
				\includegraphics[width=3.5in, height=2.6in]{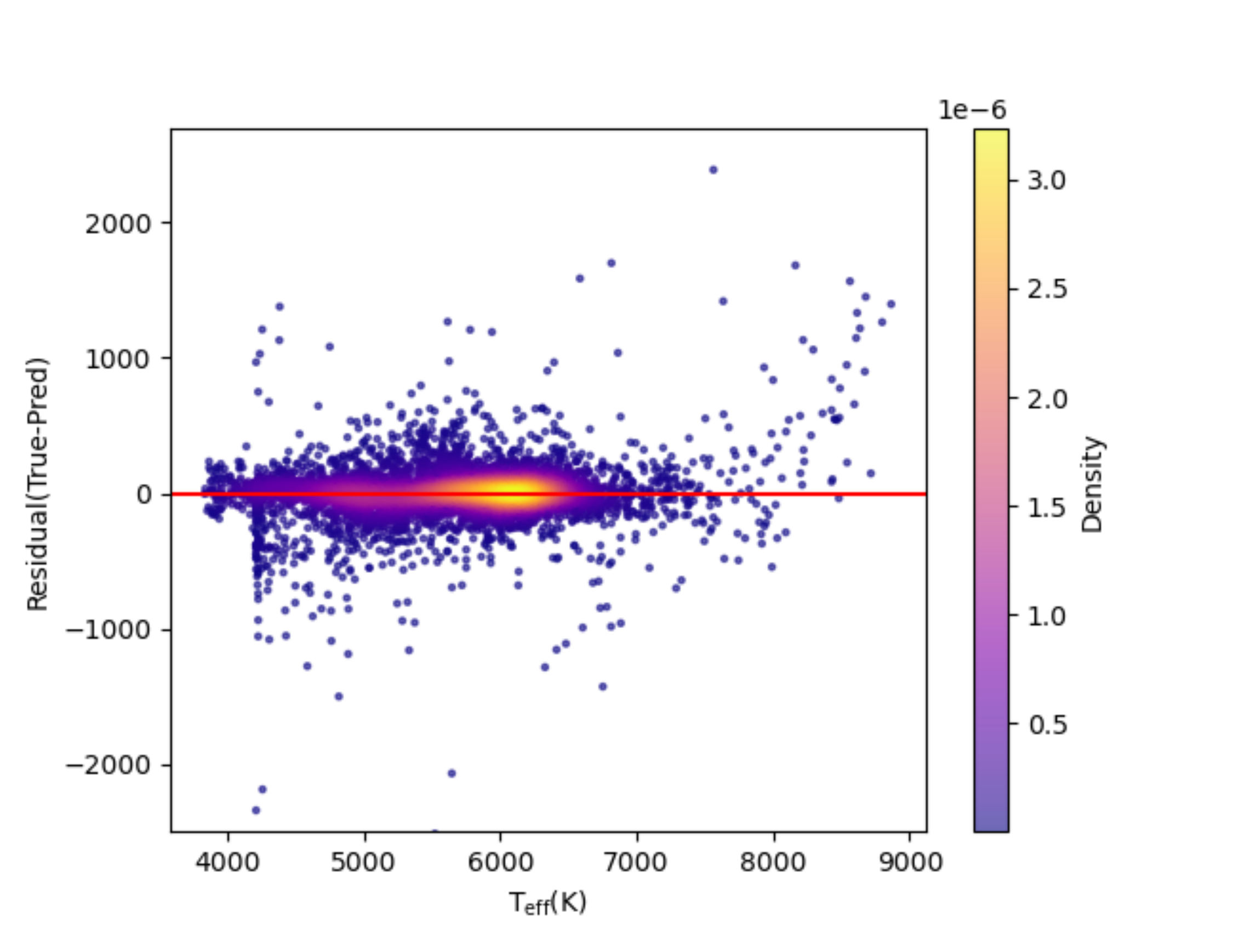}
				\centering
			\end{minipage}%
		}\hspace{-3mm}%
		\subfigure{
			\begin{minipage}[t]{0.49\linewidth}
				\centering
				\includegraphics[width=3.5in, height=2.6in]{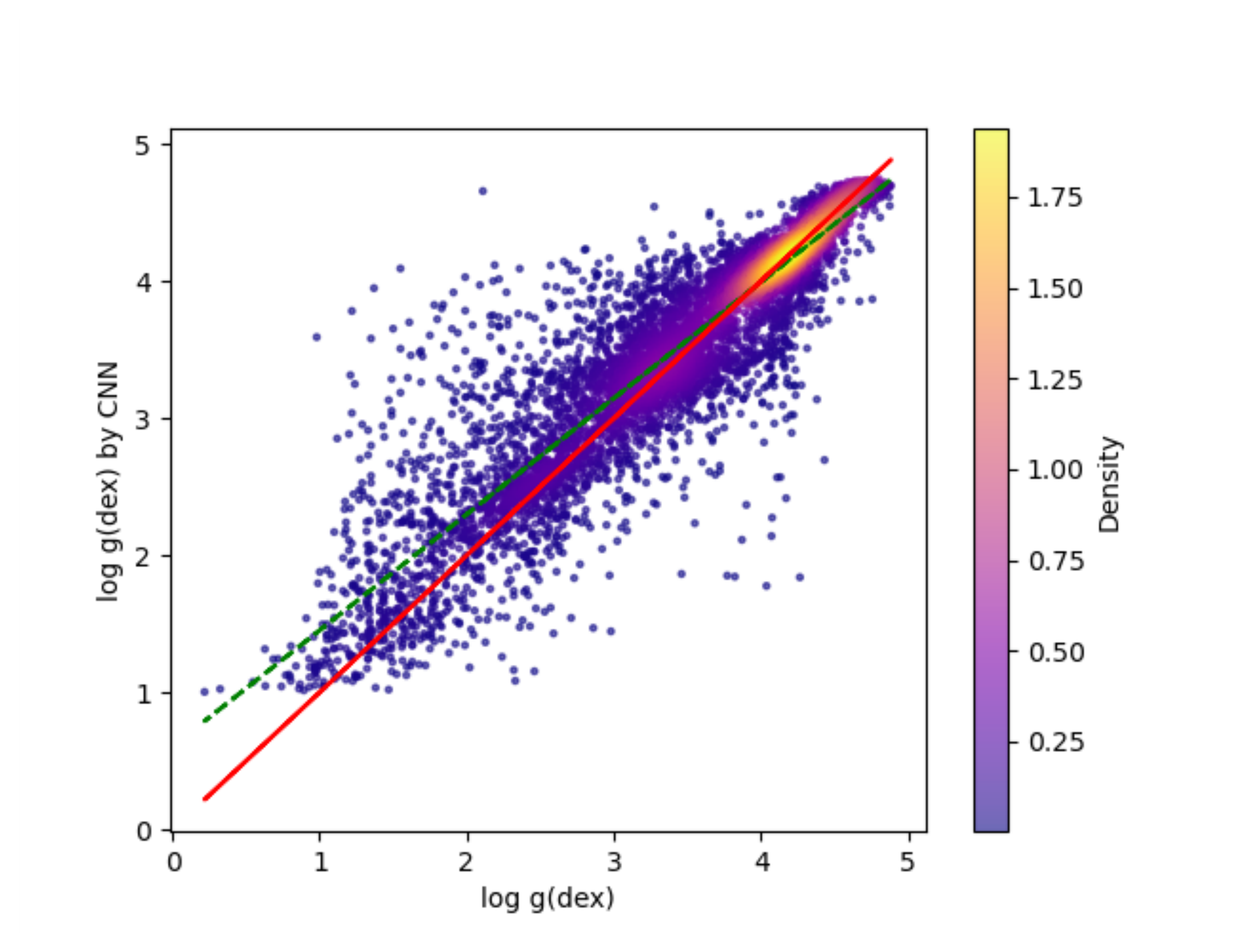}
				\centering
			\end{minipage}%
        }\hspace{-3mm}%
        \subfigure{
			\begin{minipage}[t]{0.49\linewidth}
				\centering
				\includegraphics[width=3.5in, height=2.6in]{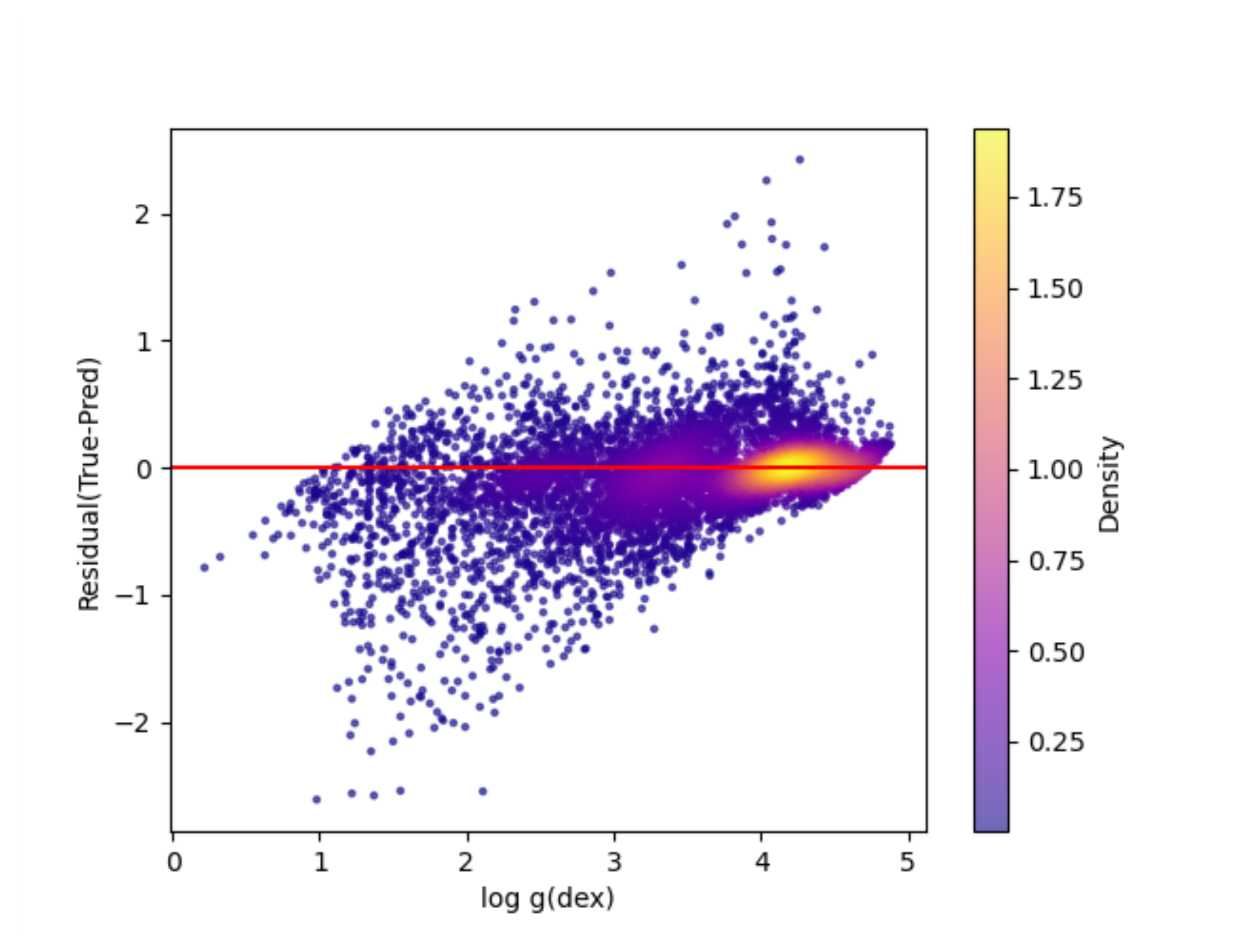}
				\centering
			\end{minipage}%
		}\hspace{-3mm}%
		\subfigure{
			\begin{minipage}[t]{0.49\linewidth}
				\centering
				\includegraphics[width=3.5in, height=2.6in]{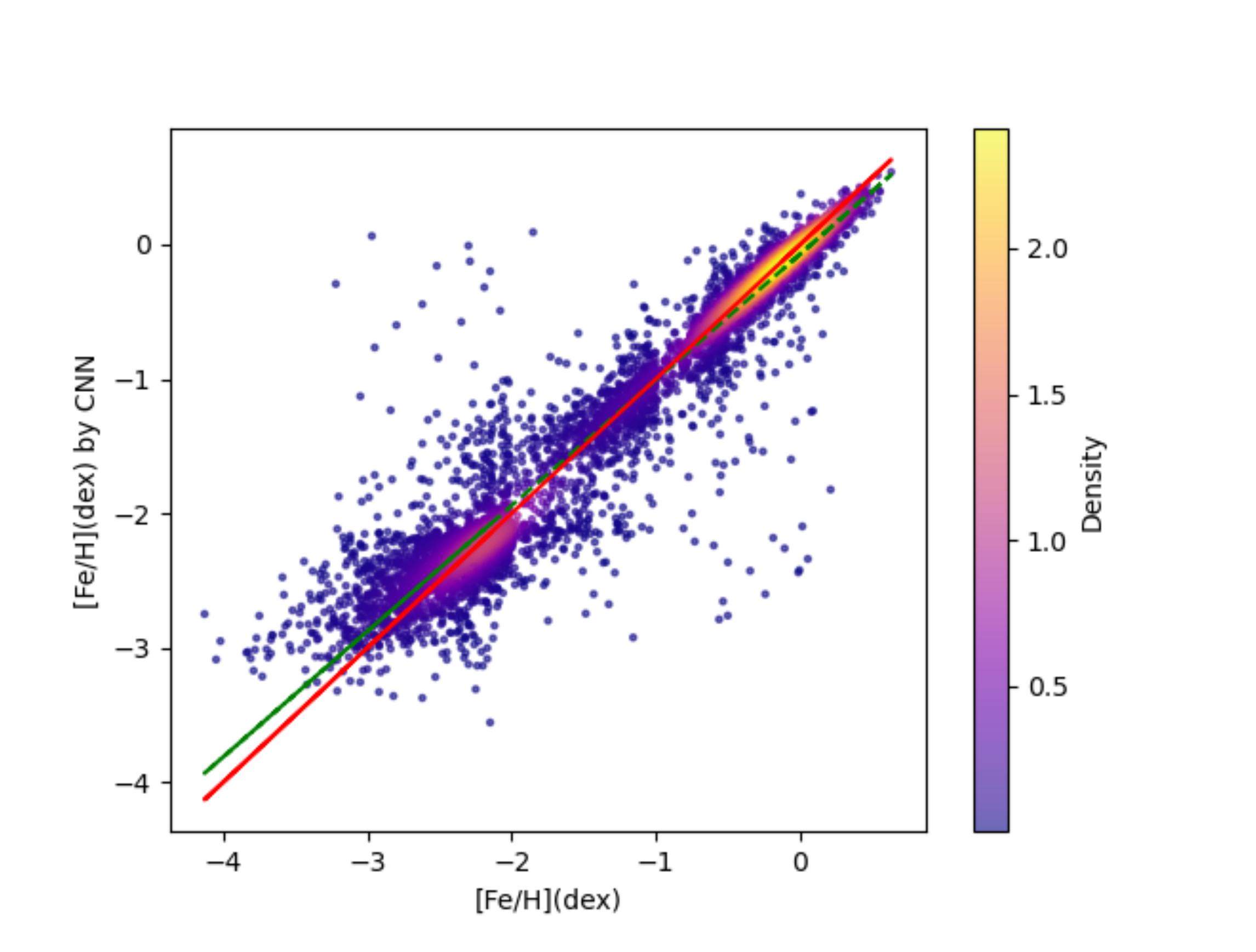}
				\centering
			\end{minipage}%
		}\hspace{-3mm}%
		\subfigure{
			\begin{minipage}[t]{0.49\linewidth}
				\centering
				\includegraphics[width=3.5in, height=2.6in]{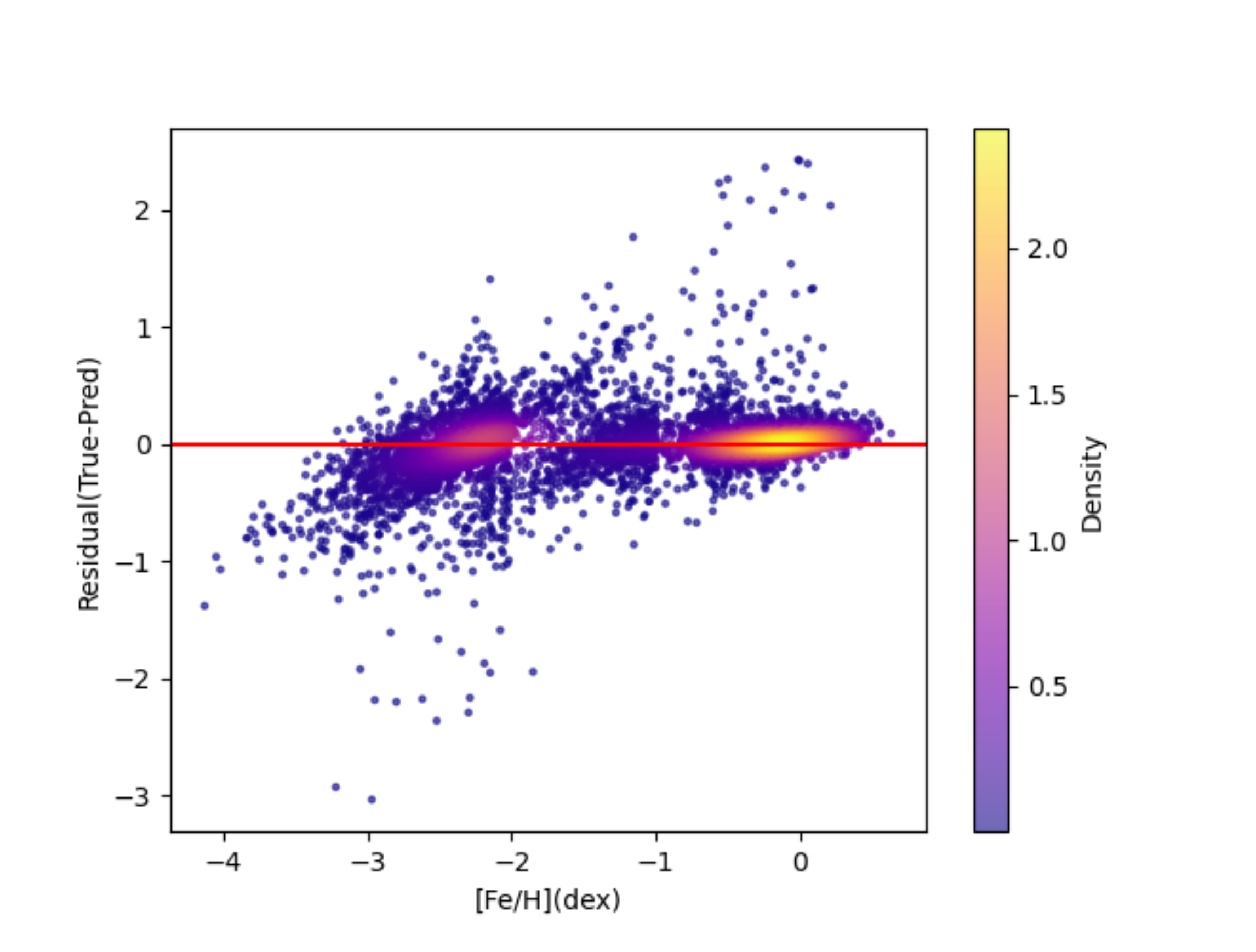}
				\centering
			\end{minipage}%
		}%
		\caption{The scatter (left) and residual (right) plots of the three fundamental atmospheric parameters on test set including 7,994  stars using the proposed CNN model.}
		\centering
		\label{fig:scatter plots}
\end{figure*}

Simultaneously, we plot the scatter density plots of the predicted and true values on the test set (see the left column of the Figure \ref{fig:scatter plots}). The green dashed line indicates the first-degree polynomial fit curve of the predicted and true values, and the red line is the image of $predicted\,value=true\,value$. The closer the green dashed line is to the red solid line, the better the prediction. From the figure, we can state that the fitting results of $T_{\rm{eff}}$ and [Fe/H] are very close to $predicted\,value=true\,value$, while the results of $\log$ g are relatively poor, which represents that the proposed CNN model has a better prediction for $T_{\rm{eff}}$ and [Fe/H], while $\log$ g is relatively more difficult to estimate. In addition, the right column of the Figure \ref{fig:scatter plots} illustrates the variation of the residuals (true value-predicted value) with respect to the true values. The red line can show us more explicitly the turbulence of the residuals around zero.

Furthermore, experiments can be conducted on 10,008 VMP stars and 16,638 common stars separately to test whether the CNN model has a significant difference in measuring the parameters of VMP stars and those of common stars. 
The VMP star dataset and the common star dataset are divided into training and test sets in the ratio of 7:3, respectively. The prediction results on the two test sets involving 3,003 VMP stars and 4,992 common stars are listed in Table \ref{tab:table2}. Briefly, the MAE values for the predicted and true values are 118.26 K for $T_{\rm{eff}}$, 0.31 dex for $\log$ g, and 0.17 dex for [Fe/H] for the VMP stars, and 75.84 K for $T_{\rm{eff}}$, 0.11 dex for $\log$ g, and 0.08 dex for [Fe/H] for the common stars. We can clearly demonstrate that the proposed CNN model is much better at deriving the parameters of common stars than VMP stars, which specifies the necessity to develop a model that can effectively measure the parameters of VMP stars.

\begin{table*}[htbp]
	\centering
	\caption{The prediction results of the three fundamental atmospheric parameters  on VMP star test set and common star test set using the proposed CNN model.}
	\label{tab:table2}
	\renewcommand\tabcolsep{10pt}
	\begin{tabular}{cccc|ccc|ccc}
		\toprule
		\multirow{2}{*} & \multicolumn{3}{c}{$T_{\rm{eff}}$(K)} & \multicolumn{3}{c}{$\log$ g(dex)} & \multicolumn{3}{c}{[Fe/H](dex)}
	 \\
		\cmidrule(r){2-4} \cmidrule(r){5-7} \cmidrule(r){8-10}  
		              &MAE     & STD      & M     &MAE  & STD   & M      &MAE  & STD  & M    \\
		\midrule
        VMP          &118.26  &173.61   &0.27  &0.31   &0.43  &-0.013  &0.17  &0.24  &-0.007\\
		COMMON         &75.84   &156.14   &7.79  &0.11  &0.18   &0.016   &0.08  &0.13  &-0.001\\
		\bottomrule
	\end{tabular}
\end{table*}

With the results obtained above, we can conclude that the proposed CNN model has good accuracy in estimating the stellar atmospheric parameters, which indicates that we can use the method for VMP star identification. First of all, precision, recall, and accuracy are defined.
\begin{enumerate}
  \item True Positive (TP): VMP stars predicted as VMP stars.
  \item True Negative (TN): Common stars predicted as common stars.
  \item False Positive (FP): Common stars predicted as VMP stars.
  \item False Negative (FN): VMP stars predicted as common stars.
\end{enumerate}
\begin{eqnarray}
Precision = \frac{TP}{TP+FP}.\\
Recall = \frac{TP}{TP+FN}.\\
Accuracy = \frac{TP+TN}{TP+FP+TN+FN}.
\end{eqnarray}

\begin{figure}[htbp]
		\centering
		\includegraphics[width=0.9\linewidth]{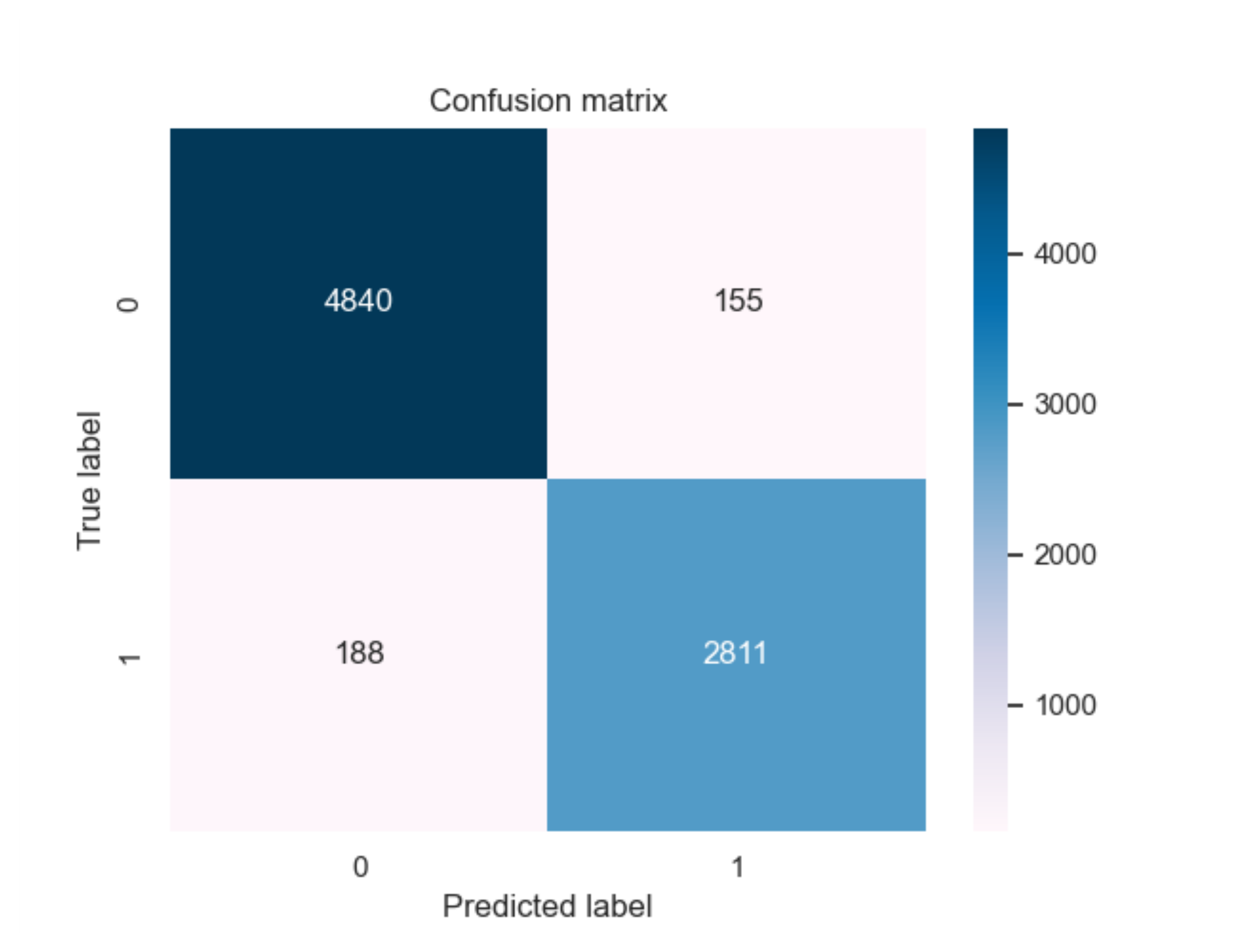}
		\caption{The confusion matrix of classifying the VMP stars on test set including 7,994 stars.}
       \label{fig:idenVMP}
\end{figure}

By analyzing the metallicity of the total dataset involving VMP and common stars, setting the label of VMP stars with [Fe/H]<-2.0 to 1 and the label of common stars with [Fe/H]>-2.0 to 0, we find 2,999 VMP stars and 4,995 common stars in the test set of 7,994 stars. The confusion matrix of the true and predicted values is shown in Figure \ref{fig:idenVMP}. Then we can calculate the precision, recall, and accuracy of the proposed CNN model for predicting VMP stars (see Table \ref{tab:tablen}). Among the 2,966 stars predicted to be VMP stars, 2,811 stars are true VMP stars, with a precision of 94.77\%; among the test set including 2,999 VMP stars, 2,811 stars are correctly predicted to be VMP stars, with a recall of 93.73\%. Overall, our CNN model is also able to classify VMP stars and common stars well, with an accuracy of 95.70\%.

\begin{table}[htbp]
  \centering
  \caption{The results of classifying the VMP stars on test set including 7,994 stars}
  \label{tab:tablen}
  \setlength{\tabcolsep}{5mm}
  \begin{tabular}{ccc}
\toprule
      Precision    & Recall  & Accuracy      \\ \midrule
        94.77\% & 93.73\%  & 95.70\% \\
\bottomrule
  \end{tabular}
\end{table}

\subsection{Estimating [C/Fe] using the VMP stars dataset} \label{subsection:c}
\begin{figure*}[htbp]
		\centering
		\subfigure{
			\begin{minipage}[t]{0.5\linewidth}
				\centering
				\includegraphics[width=3.5in, height=2.6in]{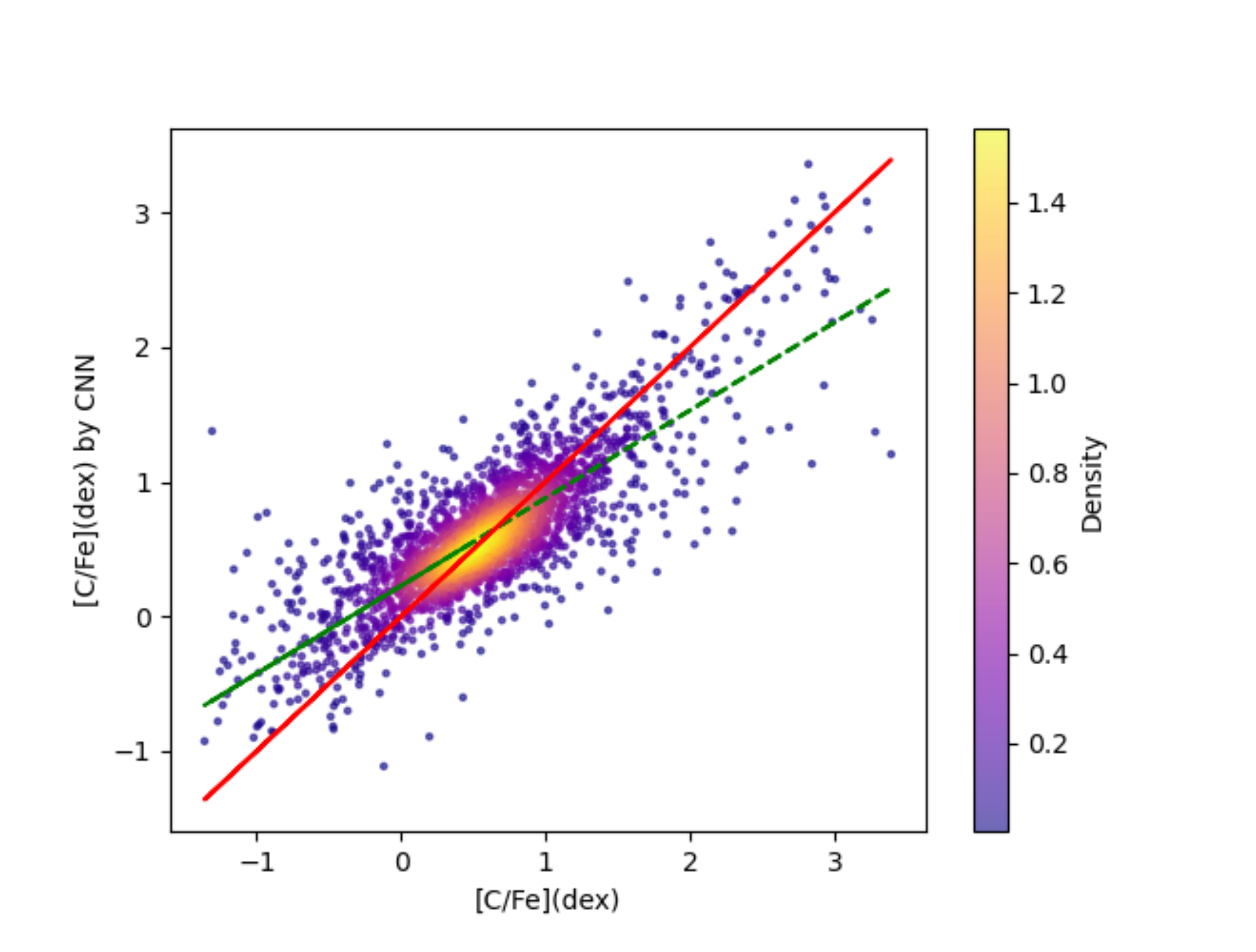}
				\centering
			\end{minipage}%
		}\hspace{-3mm}%
		\subfigure{
			\begin{minipage}[t]{0.5\linewidth}
				\centering
				\includegraphics[width=3.5in, height=2.6in]{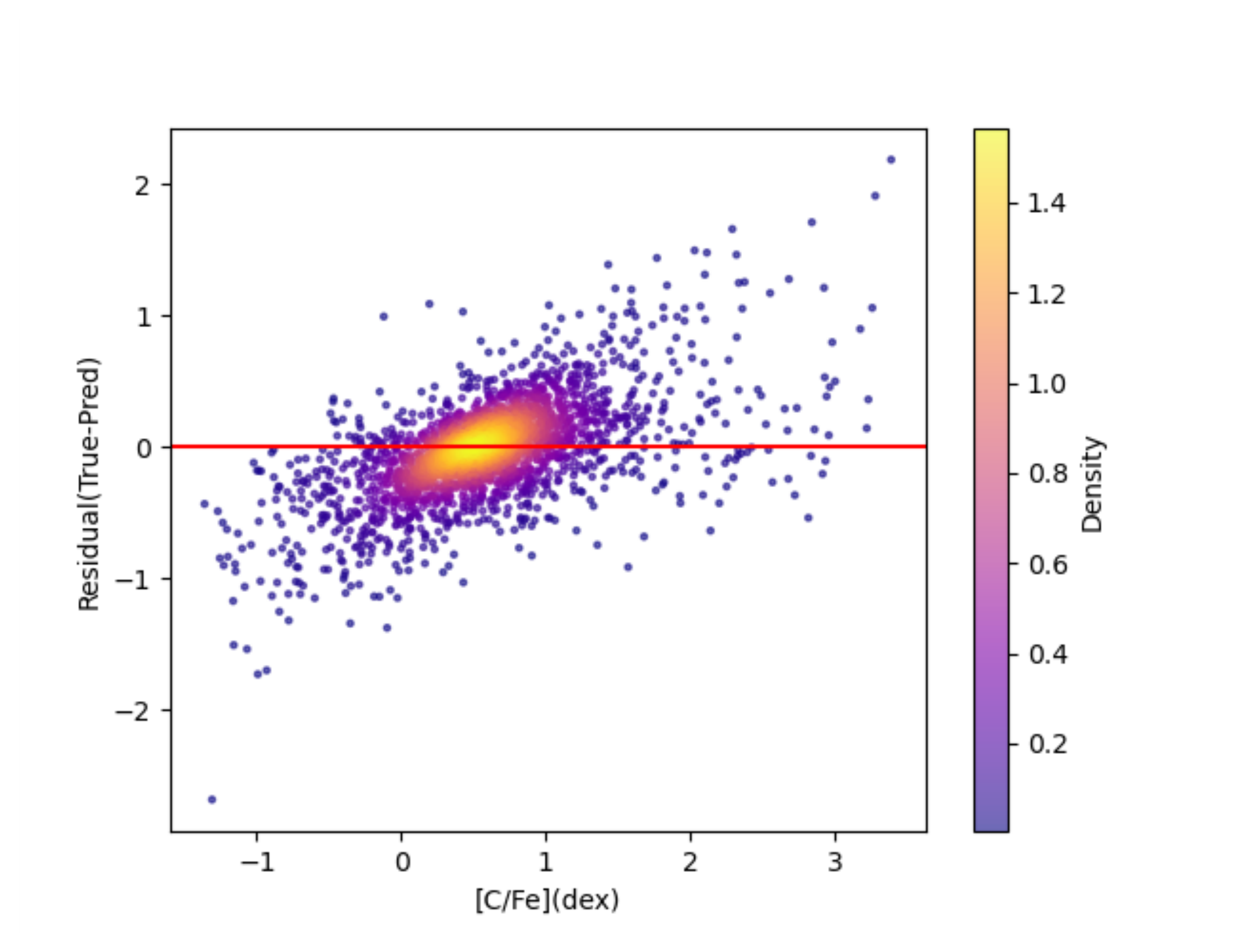}
				\centering
			\end{minipage}%
		}%
		\caption{The scatter (left) and residual (right) plots of predicting [C/Fe] on the test set including 2,436 stars using the proposed CNN model.}
		\centering
		\label{fig:cfe}
\end{figure*}

In this section, we conduct experiments using 8,117 VMP stars obtained in Section \ref{subsection:lamost}. The dataset is divided into a training set containing 5,681 stars and a test set containing 2,436 stars according to the ratio of 7:3. After 1,000 epochs of training, the best prediction results obtained on the test set are MAE=0.26 dex, STD=0.37 dex, and M=-0.01 dex. We also plot the scatter density plots of the predicted and true values of [C/Fe] and the residuals (Figure \ref{fig:cfe}). It can be seen from the figures that the proposed CNN model is also able to predict [C/Fe] well, which provides a good basis for our search for CEMP stars from VMP stars.

\subsection{Estimating stellar parameters using the MARCS synthetic spectra including 9,644 stars}\label{subsection:exp2}
\begin{figure}[htbp]
		\centering
			\begin{minipage}{\linewidth}
				
				\includegraphics[width=4in, height=3in]{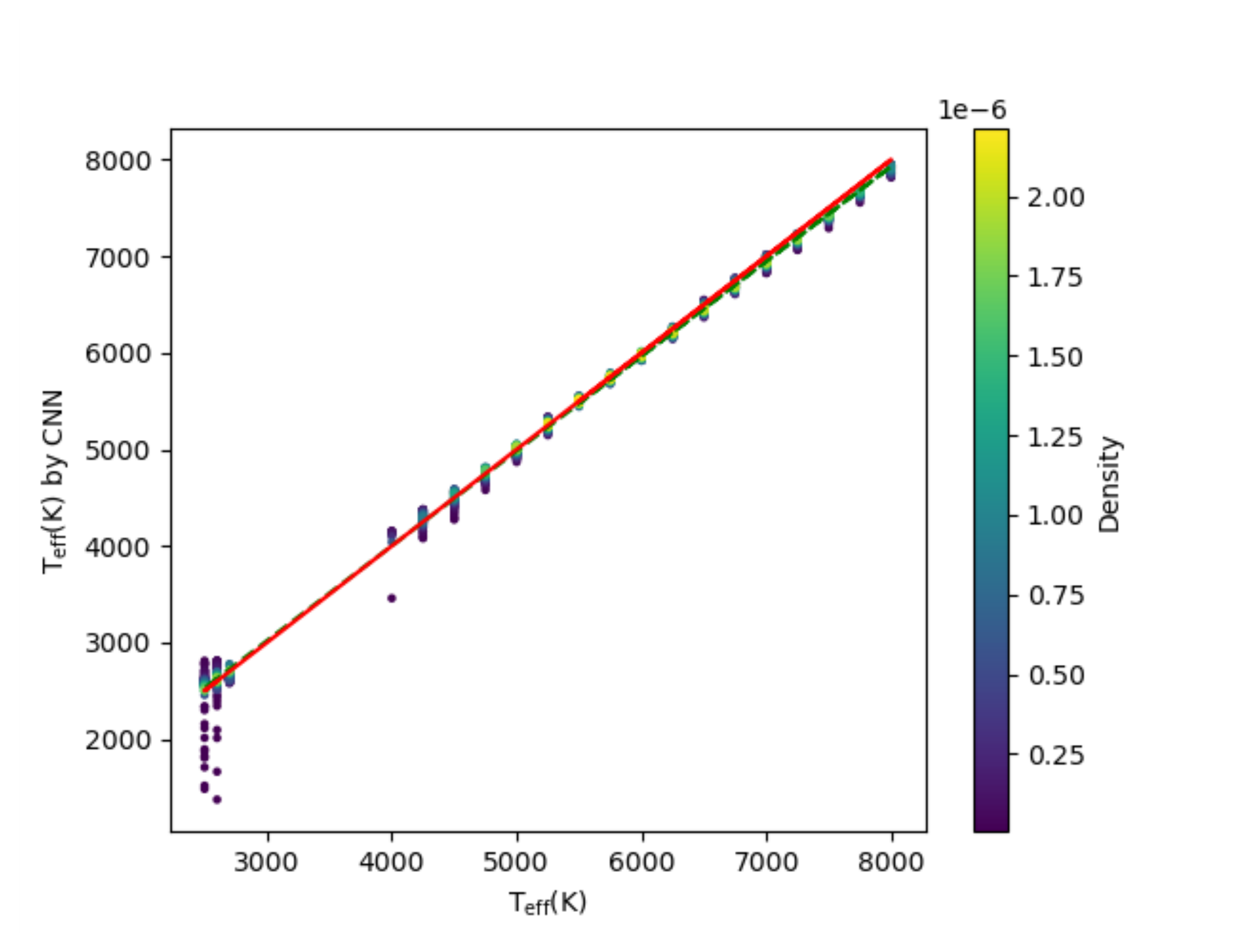}
			\end{minipage}%
		\qquad
			\begin{minipage}{\linewidth}
			
				\includegraphics[width=4in, height=3in]{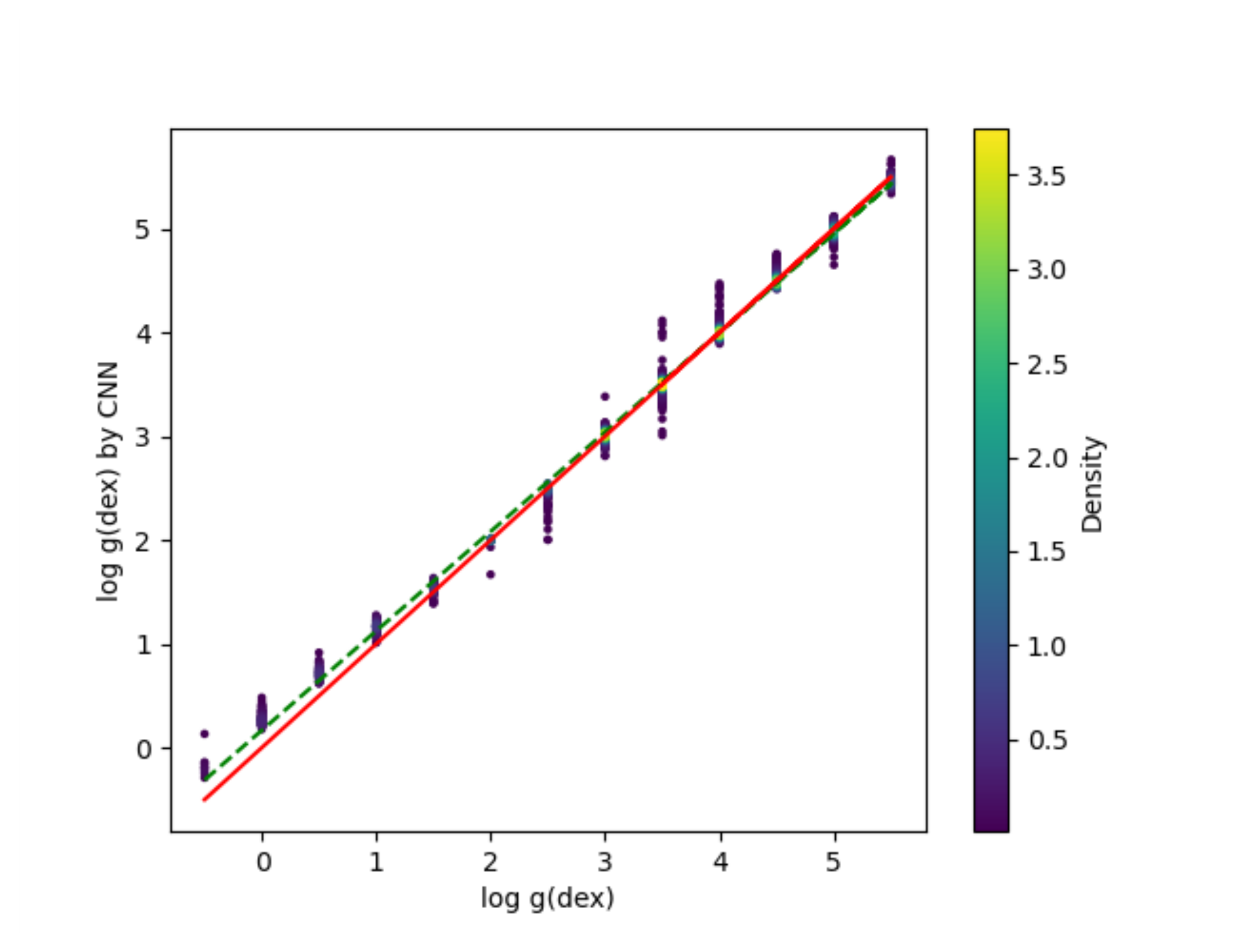}
			\end{minipage}%
		\qquad
			\begin{minipage}{\linewidth}
		
				\includegraphics[width=4in, height=3in]{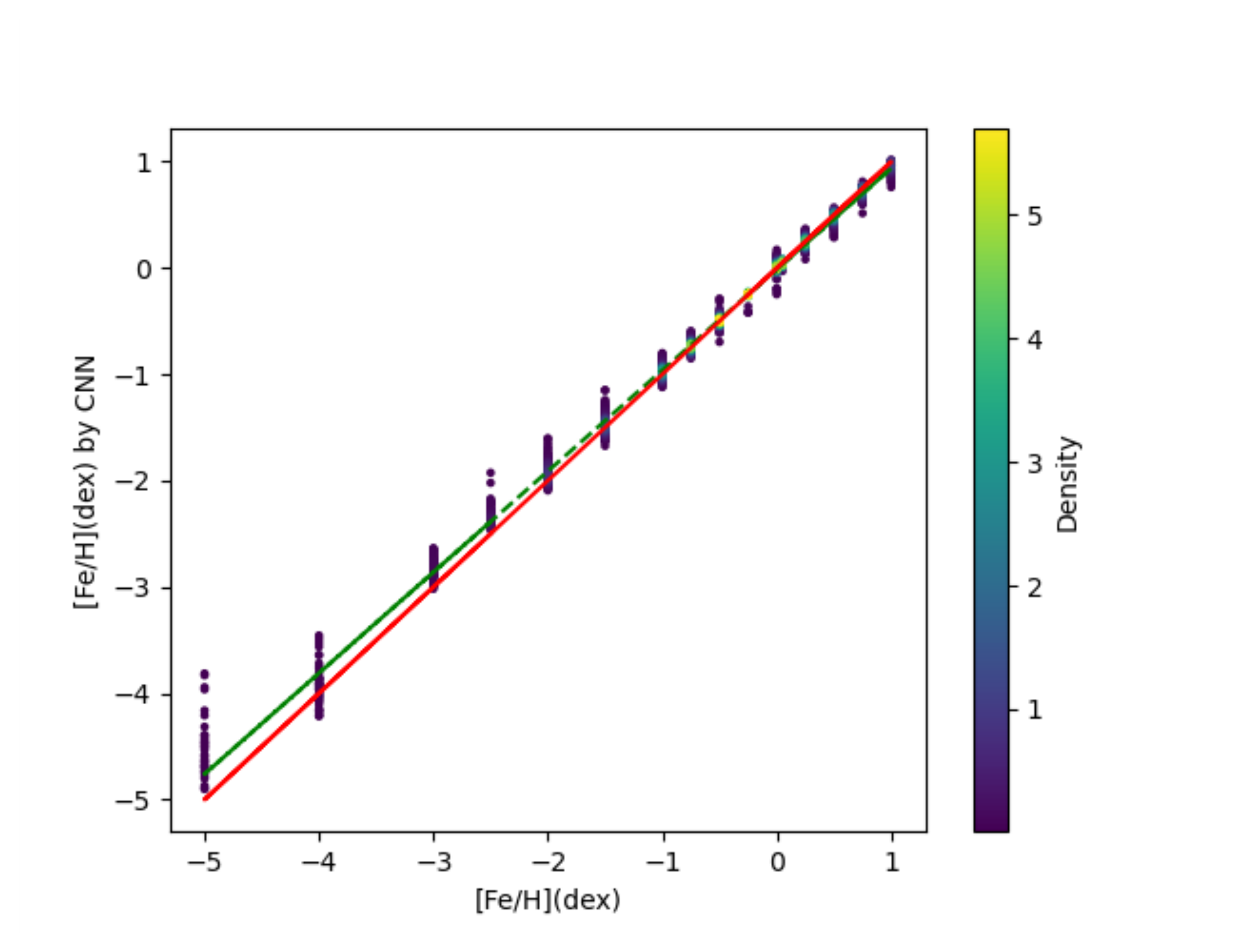}
			\end{minipage}%
		\caption{The scatter plots of the three fundamental atmospheric parameters on the MARCS test set including 2,894 stars.}
		\label{fig:fig_mock}
\end{figure}

This part displays the outcomes of predicting the stellar parameters of the 9,644 theoretical spectral data using the CNN model. The results of estimating stellar parameters using the CNN model on a test set of 2,894 stars are shown in Table \ref{tab:tablea}. For $T_{\rm{eff}}$, MAE=53.03 K, STD=80.78 K, M=12.96 K; for $\log$ g, MAE=0.056 dex, STD=0.097 dex, M=-0.002 dex; and for [Fe/H], MAE=0.047 dex, STD=0.093 dex, M=-0.004 dex. The results derived illustrate that the errors are much smaller compared to those obtained using the LAMOST data set, again proving the CNN model is valuable for estimating stellar parameters. The scatter density plots between the true and predicted values can be seen in Figure \ref{fig:fig_mock}, revealing there is only a little deviation between them.
\begin{table}[htbp]
  \centering
  \caption{The prediction results of the three fundamental atmospheric parameters on MARCS test set including 2,894 stars.}
  \label{tab:tablea}
  \setlength{\tabcolsep}{5mm}
  \begin{tabular}{cccc}
\toprule
Parameter      & MAE    & STD   & M     \\ \midrule
$T_{\rm{eff}}$ (K)      & 53.03   & 80.78  & 12.96 \\
$\log$ g(dex)       & 0.056   &0.097   &-0.002    \\
{[Fe/H]}(dex)       & 0.047   &0.093   &-0.004   \\ 
\bottomrule
  \end{tabular}
\end{table}

\section{Discussion}\label{section:discussion}
To further verify the effectiveness of the proposed CNN model, we employed Random Forest (RF) and Support Vector Machine (SVM) algorithms to make comparisons. The dataset used in the comparison experiment is the same as that used in Section \ref{subsection:exp1}, which includes a total of 26,646 spectral data and basic atmospheric parameters of the VMP stars and common stars. The training and test sets are also selected in line with the previous experiments to test whether the CNN model outperforms other algorithms. 

\begin{enumerate}
\item The RF algorithm \citep{2001MachL..45....5B}, which is a specific implementation of the bagging method, where multiple decision trees are trained and then all results are combined together. For regression problems, the prediction of the Random Forest is the average of all decision tree results. The advantage that this method can operate efficiently on large data sets and is not prone to overfitting has made it widely used in astronomical data analysis \citep{2019Spectral,2021EPJST.230.2177M}.
The RandomForestRegressor function in the Scikit-learn package in Python is imported to carry on experiments. After parameter tuning, we choose the number of decision trees as 100 and use median squared error (MSE) as the measure of quality. Setting max\_features to the number of sample features is more applicable to the regression problem. The MAE results on the test set are, 122.56 K for $T_{\rm{eff}}$, 0.30 dex for $\log$ g, and 0.26 dex for [Fe/H]. The precision, recall, and accuracy of classifying the VMP stars are 91.93\%, 75.63\%, and  88.37\%, severally.

\item The SVM approach is a binary classification model, which is essentially an optimization algorithm for solving convex quadratic programming problems. In addition to classification problems, SVM can also be applied to regression problems (SVR), which centers on finding a regression plane such that all data in a set are closest to this plane. For non-linear regression problems, SVM can introduce a kernel function that turns the problem into an approximate linear regression problem. Here we import the SVR function of the Scikit-learn library in Python, utilizing the third-degree polynomial kernel function for training, and make predictions for the test set. The MAE results on the test set are, 181.14 K for $T_{\rm{eff}}$, 0.27 dex for $\log$ g, and 0.24 dex for [Fe/H]. The precision, recall, and accuracy of classifying the VMP stars are 94.88\%, 76.69\%, and  89.70\%, respectively.
\end{enumerate}

The specific results of the comparison experiment are shown in Table \ref{tab:table3}. 
Both in terms of MAE and STD values, the estimation accuracy of CNN for stellar parameters is higher than the other two methods, and the median errors of the three methods are basically comparable. In brief, in contrast with the other two approaches, the CNN model proposed in this paper can predict the stellar parameters better.

\begin{figure}[htbp]
		\centering
		\includegraphics[width=0.9\linewidth]{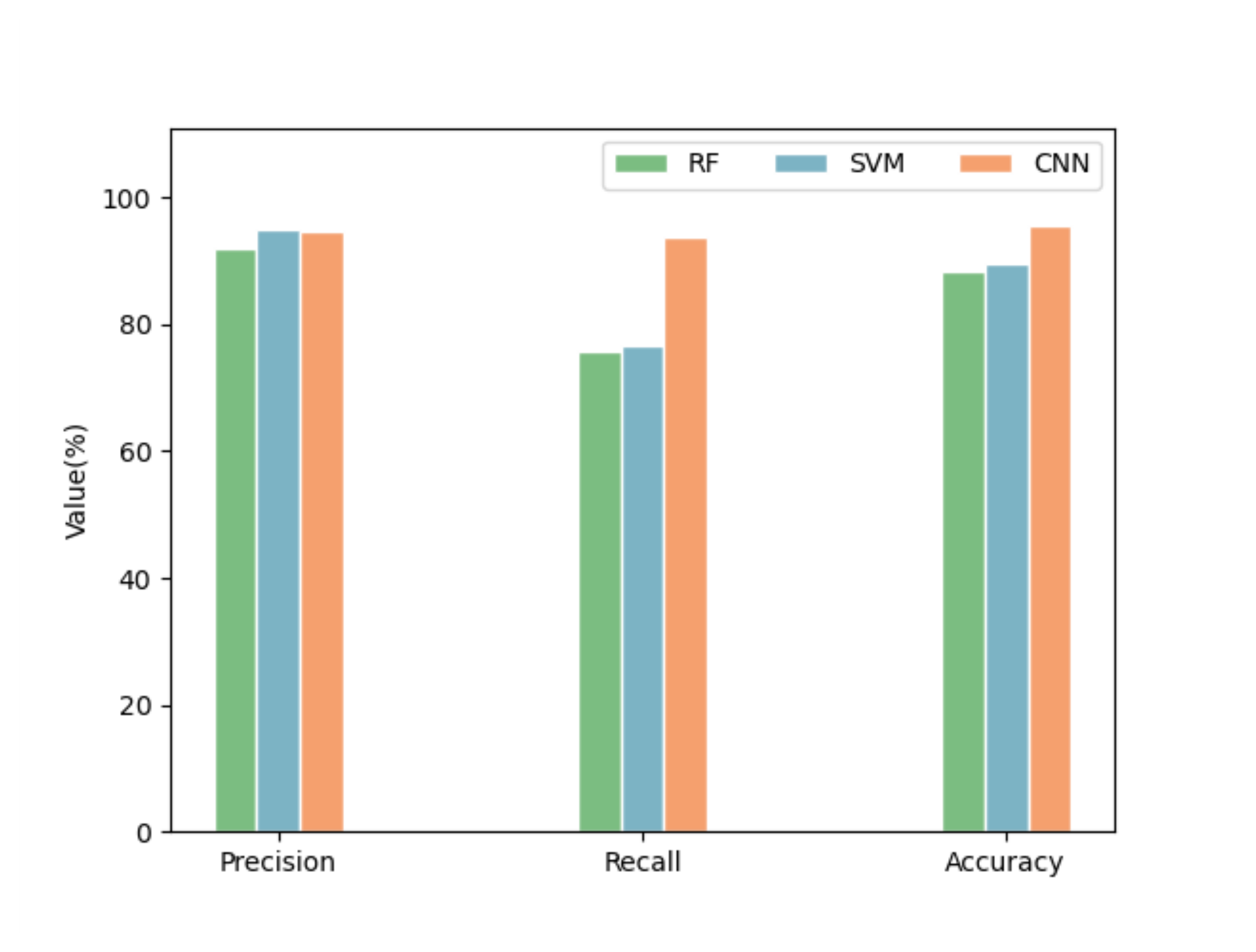}
		\caption{The bar chart of the results of identifying VMP stars on the test set including 7,994 stars using RF, SVM, and CNN.}
       \label{fig:com}
\end{figure}

We also draw a bar chart of the results of identifying VMP stars with these three methods. From Figure \ref{fig:com}, we can clearly demonstrate that although the precision of the three methods is comparable, the recall values of RF and SVM are much lower than that of CNN, which indicates that the probability of VMP stars being predicted as common stars can be greatly reduced using the proposed CNN model. In terms of accuracy, the CNN model is also able to better classify VMP stars and common stars.

\begin{table*}[htbp]
	\centering
	\caption{The prediction results of the three fundamental atmospheric parameters on the test set including 7,994  stars using RF, SVM, and CNN methods.}
	\label{tab:table3}
	\renewcommand\tabcolsep{12pt}
	\begin{tabular}{cccc|ccc|ccc}
		 \toprule
		\multirow{2}{*} & \multicolumn{3}{c}{$T_{\rm{eff}}$  (K)} & \multicolumn{3}{c}{$\log$ g  (dex)} & \multicolumn{3}{c}{[Fe/H](dex)}
	 \\
		\cmidrule(r){2-4} \cmidrule(r){5-7} \cmidrule(r){8-10}  
		          &  MAE     & STD & M   &  MAE    & STD & M  &  MAE    & STD & M \\
		\midrule
		RF      &122.56 &201.09 &4.48    &0.30 &0.45 &0.03        &0.26 &0.41 &0.02 \\
      SVR     &181.14 &302.06 &-2.31    &0.27 &0.41 &-0.003     &0.24 &0.40 &-0.002 \\
		CNN     &99.40 &183.33 &-0.49     &0.22 &0.35 &-0.02      &0.14 &0.26 &0.01\\
		\bottomrule
	\end{tabular}
\end{table*}

\section{Conclusion} \label{section:conclusion}
This paper investigates the effectiveness of the CNN model in estimating stellar parameters for low-resolution spectra ($R\sim200$) and the ability to identify VMP stars. We constructed a two-dimensional CNN model consisting of three convolutional and two fully connected layers and selected a catalog including 10,008 VMP stars and 16,638 common stars for our experiments. The resolution of these stellar spectra was reduced from $R\sim1800$ to $R\sim200$ to match the spectra obtained by CSST, and then the spectral data with 410 features could be derived by interpolation and normalization. By collapsing these one-dimensional spectra into two-dimensional matrices and feeding them into the CNN model, we can then estimate the corresponding stellar parameters. The results show that for $T_{\rm{eff}}$, MAE=99.40 K, STD=183.33 K, M=-0.49K; for $\log$ g, MAE=0.22 dex, STD=0.35 dex, M=-0.02 dex; for [Fe/H], MAE=0.14 dex, STD=0.26 dex, M=0.01 dex; and for [C/Fe], MAE=0.26 dex, STD=0.37 dex, M=-0.01 dex. Furthermore, the CNN model is slightly less capable of deriving parameters of the VMP stars compared to common stars, but it is still able to distinguish VMP stars from the test set with a precision of 94.77\%, a recall of 93.73\% and an accuracy of 95.70\%. We illustrate the advantages of the CNN model over the RF and SVM algorithms in that it can predict stellar parameters with higher accuracy and identify VMP stars better, with a recall nearly 20\% higher than the other two approaches. The efficiency of the CNN model was also tested on the MARCS synthetic spectra, and the MAE values obtained on the test set were 53.03 K for $T_{\rm{eff}}$, 0.056 dex for $\log$ g, and 0.047 dex for [Fe/H].

To sum up, the CNN model proposed in this paper can productively measure the stellar parameters of spectra with a resolution of 200 and can identify VMP stars more accurately. This work is a good foundation for future studies of a large quantity of low-resolution spectra obtained by the CSST and searching for VMP stars from them. This will not only greatly expand the VMP star candidates, but also lead to a better understanding of the evolution of the Milky Way.

\begin{acknowledgement}
This work is supported by the National Natural Science Foundation of China under grant numbers  11873037, 11603012, and 11603014 and partially supported by the Young Scholars Program of Shandong University, Weihai (2016WHWLJH09),  and the science research grants from the China manned Space Project with No CMS-CSST-2021-B05 and CMS-CSST-2021-A08.
\end{acknowledgement}

\nocite{*}
\footnotesize
\printbibliography

\end{document}